\documentclass{emulateapj}
\usepackage{graphicx}
\usepackage{epstopdf}
\usepackage{color}

\newcommand{\Mjup}{\mbox{$M_\mathrm{Jup}$}}
\newcommand{\Msun}{\mbox{$M_{\odot}$}}

\begin{document}
\shorttitle{A Substellar Companion to the Young M Dwarf 1RXS~J235133.3+312720}
\title{Planets Around Low-Mass Stars (PALMS). \\ I.  A Substellar Companion to the Young M Dwarf 1RXS~J235133.3+312720*}
\author{Brendan P. Bowler,\altaffilmark{1,2} 
Michael C. Liu,\altaffilmark{1} 
Evgenya L. Shkolnik,\altaffilmark{3}
Trent J. Dupuy,\altaffilmark{4,5}
Lucas A. Cieza,\altaffilmark{1,6}
Adam L. Kraus,\altaffilmark{1,5} 
and Motohide Tamura\altaffilmark{7}
\\ }
\email{bpbowler@ifa.hawaii.edu}

\altaffiltext{1}{Institute for Astronomy, University of Hawai`i; 2680 Woodlawn Drive, Honolulu, HI 96822, USA}
\altaffiltext{2}{Visiting Astronomer at the Infrared Telescope Facility, which is operated by the University of Hawai`i under Cooperative Agreement no. NNX-08AE38A with the National Aeronautics and Space Administration, Science Mission Directorate, Planetary Astronomy Program}
\altaffiltext{3}{Lowell Observatory, 1400 W. Mars Hill Road, Flagstaff, AZ 86001}
\altaffiltext{4}{Harvard-Smithsonian Center for Astrophysics, 60 Garden Street, Cambridge, MA 02138}
\altaffiltext{5}{Hubble Fellow}
\altaffiltext{6}{Sagan Fellow}
\altaffiltext{7}{National Astronomical Observatory of Japan, 2-21-1 Osawa, Mitaka, Tokyo 181-8588, Japan}
\altaffiltext{*}{Some of the data presented herein were obtained at the W.M. Keck Observatory, which is operated as a scientific partnership among the California Institute of Technology, the University of California and the National Aeronautics and Space Administration. The Observatory was made possible by the generous financial support of the W.M. Keck Foundation.}

\submitted{ApJ, in press (5/8/2012)}
\begin{abstract}

We report the discovery of a brown dwarf companion to the young M dwarf 1RXS~J235133.3+312720
as part of a high contrast imaging search for planets around nearby young low-mass 
stars with Keck-II/NIRC2 and Subaru/HiCIAO.  
The 2$\farcs$4 ($\sim$120~AU) pair is confirmed
to be comoving from two epochs of high resolution imaging.  
Follow-up low- and moderate-resolution near-infrared spectroscopy of 1RXS~J2351+3127~B with 
IRTF/SpeX and Keck-II/OSIRIS reveals a spectral type of L0$^{+2}_{-1}$.  The M2 primary star 1RXS~J2351+3127~A 
exhibits X-ray and UV activity levels comparable to young moving group members with
ages of $\sim$10-100~Myr.  $UVW$ kinematics based the measured radial velocity of the primary and 
the system's photometric distance (50~$\pm$~10~pc) indicate it is likely a member of the $\sim$50--150~Myr AB~Dor 
moving group.   
The near-infrared spectrum of 1RXS~J2351+3127~B does not exhibit obvious signs of youth, but 
its $H$-band morphology shows subtle hints of intermediate surface gravity.
The spectrum is also an excellent match to the $\sim$200~Myr M9 brown dwarf LP~944-20.
Assuming an age of 50--150~Myr, evolutionary models imply a mass 
of 32~$\pm$~6~\Mjup \ for the companion, making 1RXS~J2351+3127~B 
the second lowest-mass member of the AB~Dor moving group after the L4 companion CD--35~2722~B and one of the few benchmark 
brown dwarfs known at young ages.

\end{abstract}
\keywords{stars: individual (1RXS~J235133.3+312720) --- stars: low-mass, brown dwarfs}

\section{Introduction}{\label{sec:intro}}

M dwarfs are the most abundant denizens of our galaxy and because of their sheer numbers are probably the most 
common sites of planet formation (\citealt{Lada:2006p14060}).  
They account for over 70\% of stellar systems in the solar neighborhood  (\citealt{Henry:1997p22864})  and make
up about half of the baryonic mass of our galaxy (\citealt{Henry:2004p23360}).  
In addition, the single star fraction--- a crucial statistic 
for giant planet formation (e.g., \citealt{Kraus:2012p23303})--- decreases from $\sim$60-70\% for M dwarfs (\citealt{Fischer:1992p18426}; \citealt{Bergfors:2010p23053}) 
to $\sim$54\% for solar-type stars (\citealt{Duquennoy:1991p18465}; \citealt{Raghavan:2010p23127}) to near 0\% 
for  the most massive stars (\citealt{Preibisch:1999p23128}), further separating 
M dwarfs from AFGK stars as the most numerous potential planet hosts of all the stellar classes (\citealt{Lada:2006p14060}).

Despite the prevalence of M dwarfs in our galaxy, the population of giant planets orbiting low-mass stars
remains poorly understood.  At small separations ($\lesssim$3~AU), radial velocity surveys
have found that the frequency of giant planets   
decreases with diminishing stellar host mass (\citealt{Endl:2006p19580}; \citealt{Johnson:2007p169}; 
\citealt{Johnson:2010p20950}; \citealt{Bonfils:2011p22962}), a correlation that is consistent
with the core accretion model of giant planet formation (e.g., \citealt{Kennedy:2008p18349}).
At wide separations ($>$5~AU), however,
there are few observational constraints on the population of gas giants.
The statistics from microlensing surveys indicate that a large population of planets exists 
beyond the snow line of low-mass stars (\citealt{Gould:2006p23481}; \citealt{Sumi:2010p20095}; 
\citealt{Gould:2010p23021}; \citealt{Sumi:2011p22269}).  Recently, \citet{Cassan:2012p23478} 
measured the occurrence rate of 0.3--10~\Mjup \ planets between
0.5--10~AU to be 17$^{+6}_{-9}$\%, with the frequency increasing for lower planet masses.
Unfortunately, the host stars from this sample span a large mass range 
(68\% have masses between 0.14 and 1.0 \Msun) and their metallicities are 
unconstrained, so it is unclear how these parameters might influence planet formation
at these separations.

Direct imaging offers another approach to study planetary systems at wide separations.
This method has several advantages over other planet-finding techniques: it enables
detailed studies of regions $\gtrsim$10~AU, and follow-up photometry and 
spectroscopy of discoveries can be used to study the atmospheres of young giant planets.
Yet M dwarfs are conspicuously rare targets of direct imaging surveys due 
in part to a dearth of known low-mass members of nearby young moving groups 
(e.g., \citealt{Shkolnik:2009p19565}; \citealt{Shkolnik:2011p21923}; \citealt{Schlieder:2012p23477}).  
Most deep adaptive optics imaging programs have therefore 
focused on the more abundant solar- and high-mass members of these groups, which has lead to
increasingly tighter statistical constraints on the population of massive planets on
wide orbits around FGK stars (\citealt{Lafreniere:2007p17991}; \citealt{Biller:2007p19401}; 
\citealt{Nielsen:2008p18024}; \citealt{Nielsen:2010p20955}; \citealt{Chauvin:2010p20082}).
M dwarfs are therefore the neglected majority: they are the most
common stellar type  but our understanding of giant planet formation is the 
weakest in this stellar mass regime.

To address these disproportionate statistics we are carrying out the Planets Around Low-Mass Stars 
(PALMS) survey, a high contrast adaptive
optics (AO) imaging search for giant planets around nearby ($\lesssim$30~pc) 
young ($\lesssim$300~Myr) M dwarfs using the Keck-II and Subaru telescopes.
Our goals are to find young giant planets and brown dwarf companions to study
the atmospheres of these rare low-gravity objects and 
measure the frequency and mass-period distributions of gas giants 
orbiting M dwarfs.  Our sample of $\sim$70 northern targets 
originates from ongoing searches for nearby young M dwarfs using the $ROSAT$ and $GALEX$
all-sky surveys (\citealt{Shkolnik:2009p19565}; \citealt{Shkolnik:2011p21923}).
High resolution optical spectroscopy has been used to rule out spectroscopic binarity, identify
spectroscopic indicators of youth like Li absorption and gravity-sensitive features, and
measure radial velocities.  Many of the targets in our sample have been kinematically tied to 
young moving groups (YMGs) with ages between 10 and 100~Myr (Shkolnik et al. 2012, submitted).
Recently, several studies have found that M dwarfs hosting close-in giant planets are preferentially 
metal-rich (e.g., \citealt{Johnson:2009p18908}; \citealt{RojasAyala:2010p21649}; \citealt{Terrien:2012p23659}).  
On average, our sample of young stars is expected to be slightly 
metal-rich as a result of galactic chemical enrichment over time. 
Our PALMS survey therefore complements ongoing radial velocity planet searches
that are uncovering gas giants around metal-rich field M dwarfs.  Finally, we note that only a handful of giant planets
have been detected around M dwarfs (7 with masses  $>$ 1 \Mjup \ from radial velocity and microlensing 
searches; see the compilation in \citealt{Bonfils:2011p22962}), so even a single discovery from 
our survey is significant.

A handful of planets have now been directly imaged 
and many more are expected
to be discovered with the next generation of specialized  planet-finding 
instruments (i.e., Gemini Planet Imager, \citealt{Macintosh:2006p23015}; HiCIAO+SCExAO, 
\citealt{Martinache:2009p23016}; P1640+PALM-3000, \citealt{Hinkley:2011p23019}; 
VLT-SPHERE, \citealt{Beuzit:2008p23018}).  
Many of our expectations about the atmospheric properties of giant planets
are shaped by detailed studies of their higher-mass analogs, the brown dwarfs.  With similar
radii and effective temperatures to young gas giants, brown dwarfs provide 
empirical spectral sequences
across a range of gravities and metallicities.  They also serve as tests of substellar 
atmospheric and evolutionary models, which are the same
models used to infer the physical properties of giant planets.  
The best calibrators are the rare class of benchmark brown dwarfs 
with known ages and metallicities which can be found as members of coeval clusters 
(e.g., \citealt{Lodieu:2008p8698}; \citealt{Rice:2010p20833}) 
or companions to well-characterized stars (e.g, \citealt{Liu:2002p18017}; 
\citealt{Luhman:2007p10341}; \citealt{Dupuy:2009p15627};  \citealt{Wahhaj:2011p22103}; 
\citealt{Crepp:2011p23186}).

\begin{figure}
  \resizebox{\textwidth}{!}{\includegraphics{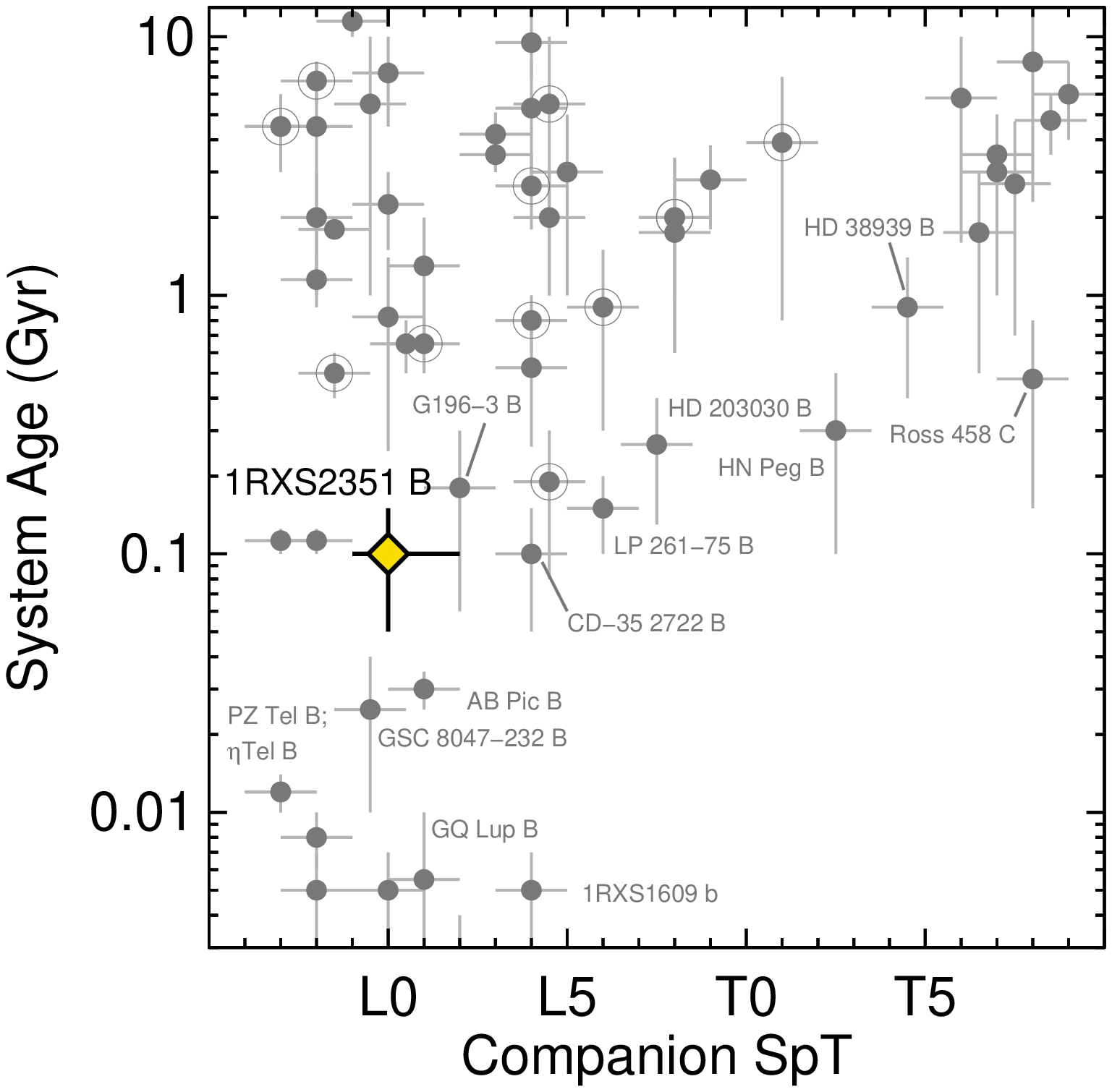}}
  \caption{Census of benchmark ultracool companions compiled from the literature.  The sample is limited to main-sequence primaries
  with spectral types earlier than M5 and systems with well-constrained ages.  Companions resolved 
  into tight binaries are indicated with open circles and some of the younger/more recent discoveries are labeled.  
  The HR~8799 planets are excluded because of their peculiar spectral and photometric properties 
  (e.g., \citealt{Bowler:2010p21344}; \citealt{Barman:2011p22098}).
   In addition to 1RXS~J2351+3127~AB only a handful other systems with ages of $\sim$50-300~Myr are known.  
  A number of interesting features are apparent from this figure.  There is a dearth of mid-L to T dwarf companions at 
  young ages ($\lesssim$200~Myr), which corresponds to objects near and below the deuterium-burning limit. 
  At old ages ($\gtrsim$7~Gyr) no late-L or early-T dwarf companions are known.  While this
  could be a selection effect, it may also be showing the luminosity/temperature
  gap differentiating the lowest-mass stars, which have nearly constant temperatures at old ages, from brown dwarfs,
  which cool over time.  Note that the compilation is incomplete for late-M dwarfs, especially at old ages.\label{fig:benchmarks} } 
\end{figure}

Here we present the first discovery from our PALMS survey: a substellar companion to the 
young M2 dwarf 1RXS~J235133.3+312720 (hereinafter 1RXS~J2351+3127~A).  
This young star was first identified by \citet{Riaz:2006p20030} and 
\citet{Shkolnik:2009p19565} from its large fractional X-ray flux.
Shkolnik et al. (2012, submitted) and \citet{Schlieder:2012p23477} independently find
it to be a likely member of the AB~Dor YMG based on its high energy emission and kinematics.
Our discovery of a substellar companion to 1RXS~J2351+3127~A makes it an important 
benchmark system at a little-studied age of brown dwarf evolution (Figure~\ref{fig:benchmarks}).

\section{Observations}{\label{sec:obs}}

\subsection{PALMS Observing Strategy}{\label{sec:palmsobs}}

The overall strategy of our survey consists of vetting close visual binaries from our 
sample and obtaining deep imaging of single young M dwarfs.  Binaries are removed
for several reasons: (1) moderate-separation binaries ($\lesssim$40~AU) are likely disruptive to giant planet 
formation as a result of rapid disk dispersal (\citealt{Kraus:2012p23303}); (2) for giant planets forming in 
circumbinary disks, 
very tight binaries can be approximated as single point masses made of the total mass of the binary pair, and 
therefore cannot be included in our final statistical analysis which is focused on low stellar masses; and 
(3) wavefront correction in adaptive optics systems is generally optimized for single point sources, 
so binaries will tend to reduce image quality and lower Strehl ratios.  Our deep coronagraphic imaging of each target
is conducted in angular differential imaging (ADI) mode (e.g., \citealt{Liu:2004p17588}; \citealt{Marois:2006p18009}); 
these data will be discussed in a future publication.

\subsection{Keck-II/NIRC2 NGS AO Imaging}{\label{sec:obsnirc2}}

\begin{figure}
  \vskip -1 cm
  \resizebox{\textwidth}{!}{\includegraphics{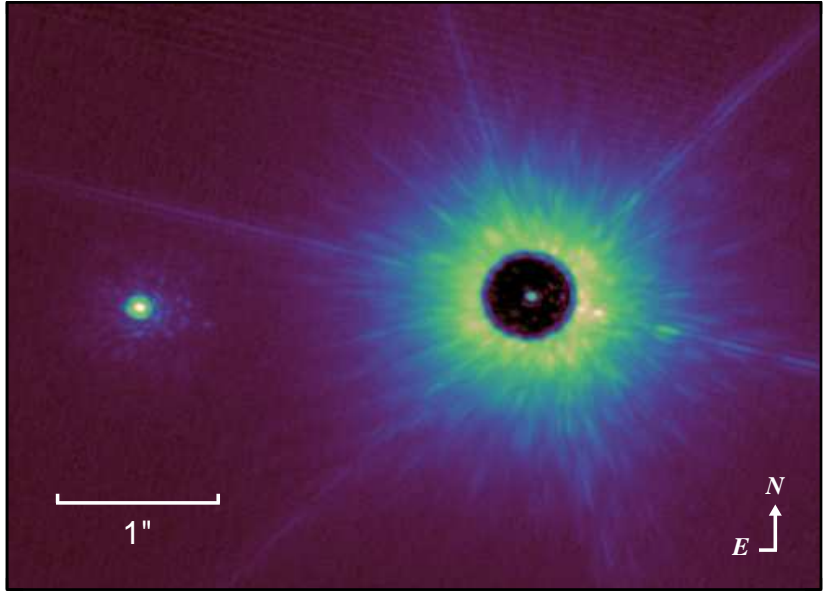}}
  \caption{Keck/NIRC2 coadded $H$-band image of 1RXS~J2351+3127~AB.  The primary is positioned behind 
  the 0$\farcs$6 diameter
  translucent coronagraph with the companion located to its east.  Both objects appear to be single down to the
  diffraction limit of Keck ($\sim$40~mas).  The image is displayed with an asinh stretch (\citealt{Lupton:2004p20516})
  and the ``cubehelix" color scheme of \citet{Green:2011p23338}.
  \label{fig:nirc2} } 
\end{figure}

We imaged 1RXS~J2351+3127~AB with Keck-II/Near Infrared Camera~2 (NIRC2) coupled with 
natural guide star (NGS) adaptive optics (\citealt{Wizinowich:2000p21634})
on 21 June 2011 UT and 15 November 2011 UT.  The narrow camera setting
was used for both epochs, resulting in a field of view of 10$\farcs$2$\times$10$\farcs$2.
Our June 2011 data were obtained at an airmass of 1.17 with the MKO $H$-band filter 
(\citealt{Simons:2002p20490}; \citealt{Tokunaga:2005p18542}).  We first acquired 
short unsaturated images
of the primary (reading out the central 192$\times$248 pixels instead of the full 
1024$\times$1024 array) for photometric calibration and to check for binarity.  
We then obtained five 10 sec images using the full array
with 1RXS~J2351+3127~A placed behind the 0$\farcs$6 diameter translucent focal plane mask; 
1RXS~J2351+3127~B is clearly visible in individual frames (Figure~\ref{fig:nirc2}).  Conditions were
photometric and the Differential Imaging Motion Monitor on CFHT reported 0$\farcs$4 seeing during
the observations.
Second epoch imaging in November 2011 was performed in the $K'$-band filter 
at an airmass of 1.05.
Nine short (0.2~sec) non-coronagraphic frames of 1RXS~J2351+3127~AB were acquired during  
excellent ($\sim$0$\farcs$5) conditions.  Our imaging data are summarized in Table~\ref{tab:obs}.

Cosmic ray and bad pixel removal, dark subtraction, and
flat fielding were performed on each image.  The coronagraphic frames suffer from dust features 
present on the focal plane slide which contains the occulting spot.
This slide may
not return to exactly the same position with each new setup, so we created two flat frames   
to remove normal detector and optical inhomogeneities as well as residual dust 
features from this nonstatic optical component.
We obtained dome flats with and without the mask, then
generated a ``coronagraph" flat by dividing the 
normal dome flat.   After masking out the 
occulting spot from the coronagraph flat (that is, the region centered on the coronagraph spot was
set to unity), we matched the flat to the image through cross
correlation, then divided it into the image.  
Optical distortions were corrected using the distortion solution made available
by the Keck Observatory, which was developed by B. Cameron and
is accurate to $\sim$0.2-0.3 pix across the entire image (\citealt{Yelda:2010p21662}).

Our $H$-band coronagraphic data preceeded deep angular differential
imaging (which will be presented in a future publication) so the image rotator
was turned off for the observations (i.e., the rotator was in``vertical angle mode''), 
causing 1RXS~J2351+3127~B to rotate about the
primary by 0.34$^{\circ}$ during the short sequence.  The images were registered by 
fitting a 2D elliptical Gaussian
to the primary star behind the coronagraph and derotated to a common position angle 
using a cubic convolution 
interpolation.  North alignment was performed using keywords stored in the FITS header, 
taking into account the offset between the AO and NIRC2 detector (+0.7$^{\circ}$) and 
the sky orientation on the detector of +0.252~$\pm$~0.009$^{\circ}$ derived by \citet{Yelda:2010p21662}.

Astrometry was measured for each image using centroid positions of 1RXS~J2351+3127~A and B
and the NIRC2 plate scale of 9.952$\pm$0.002 mas pix$^{-1}$ from \citet{Yelda:2010p21662}.
Errors in the separation and position angle (PA) were computed from Monte Carlo realizations taking into
account uncertainty in the North angle alignment, plate scale, centroid position (assumed to be 0.1 pix),
and distortion solution (0.3 pix).  The adopted values for each epoch are weighted averages
from the individual frames. (The uncertainties in separation and PA from individual images are 
roughly 4~mas and 0.1$^{\circ}$, respectively.)
Our measurements are summarized in Table~\ref{tab:astrometry}.

We computed flux ratios for our $H$-band data using the short images of the primary
and the longer 10 sec coronagraphic frames where the companion is visible.  We performed aperture 
photometry at the centroided positions of both components using an aperture
radius of 10 pix and annular sky subtraction, arriving at an $H$-band flux ratio 5.68 $\pm$ 0.04 mag 
and a $K'$-band flux ratio of 5.04 $\pm$ 0.05 mag.

\subsection{IRTF/SpeX Prism Near-Infrared Spectroscopy}\label{sec:prismobs}

We obtained a low-resolution 0.8--2.5~$\mu$m spectrum of 1RXS~J2351+3127~B with IRTF/SpeX 
(\citealt{Rayner:2003p2588}) in prism
mode on 14 October 2011 UT.  The 0$\farcs$3 slit was used (2 pixels per resolution element), 
resulting in an average resolving power ($R \equiv \lambda /  \Delta \lambda $) of $\sim$250 
across the spectrum.  Atmospheric conditions were good during the
observations (DIMM on CFHT reported 0$\farcs$5 seeing) with some light cirrus.  The slit was oriented 
perpendicular to the binary PA; although this differed from the parallactic angle, differential chromatic refraction was  
negligible because of the low airmass (sec$z$=1.1).  A total of 12 min of data were obtained
by nodding along the slit in an ABBA pattern (Table~\ref{tab:obs}).  Immediately afterward we observed the A0V standard
HD~222749 for telluric correction at a similar airmass (sec$z$=1.2) and position on the sky.  Internal
flats and arc frames were taken for flat fielding and wavelength calibration.  The spectra were extracted,
median-combined, and corrected for telluric features using the IDL package Spextool 
(\citealt{Cushing:2004p501}; \citealt{Vacca:2003p497}).  The median S/N per pixel 
between 0.8--2.5~$\mu$m is 74 and reaches over 130 in $J$ band.

The spectrum of 1RXS~J2351+3127~B suffers from significant contamination from the primary at 
$\lambda \lesssim$1.2~$\mu$m,
which was evident in the pair-subtracted images.  However, the smaller PSF FWHM at longer wavelengths
resulted in a better separation at $H$- and $K$-bands, so 
contamination should be negligible in those regions.  We flux calibrated the spectrum using the 
$K$-band photometry from Keck and the conversion to the MKO system  described in Section~\ref{sec:dist}.

\subsection{IRTF/SpeX SXD Near-Infrared Spectroscopy}

We observed 1RXS~J2351+3127~AB with the IRTF/SpeX spectrograph in short cross-dispersed mode (SXD) on 
2 Dec 2011 UT.  Conditions were excellent with DIMM on CFHT reporting 0$\farcs$4 seeing.
We used the 0$\farcs$5 slit ($R \sim 1200$) oriented at the binary position angle so that both the primary 
and companion were observed.  The observations were taken in an ABBA pattern over an airmass
range from 1.02 to 1.06.  We obtained 24 120-sec exposures, resulting in a total
on-source integration time of 48 min.  The A0V star 7~Tri was then targeted for telluric correction, and calibration 
frames were taken at the same telescope position.  

The data were reduced with Spextool.   It was clear from the collapsed spatial profiles of each order that the wing of the 
point spread function 
from the primary overlapped with the companion.  Like in the prism data, the system was better separated at $H$ and $K$
because of the smaller FWHM, but contamination became progressively worse at $\lesssim 1.1$~$\mu$m.  
As a result, we processed the spectra of the
companion without 
optimal extraction and sky subtraction, and we limited the extraction to orders 3, 4, and 5 ($K$, $H$, and $J$ bands, respectively, or 1.1--2.5~$\mu$m).  
The SXD spectrum appears to be slightly redder than the prism spectrum, which may indicate there is less
contamination.  (The synthetic $J$--$K$ (MKO) colors of the SXD and prism spectra are 1.26~mag and 1.17~mag.)
The 4th order of the companion fell on a region of the detector where a faint ghost exists.  
Consequently, a large artifact was present in the reduced spectrum between 1.71--1.75~$\mu$m, so we removed
this spectral region.  The extraction of the primary star and the standard were performed 
the usual way with Spextool using optimal extraction and background subtraction.
The median S/N per pixel of the companion spectrum is 26 and is over 40 in $H$ and $K$.

\subsection{IRTF/SpeX Guider Camera $YJHK$ Imaging}

\begin{figure}  
\begin{center}
  \resizebox{5in}{!}{\includegraphics{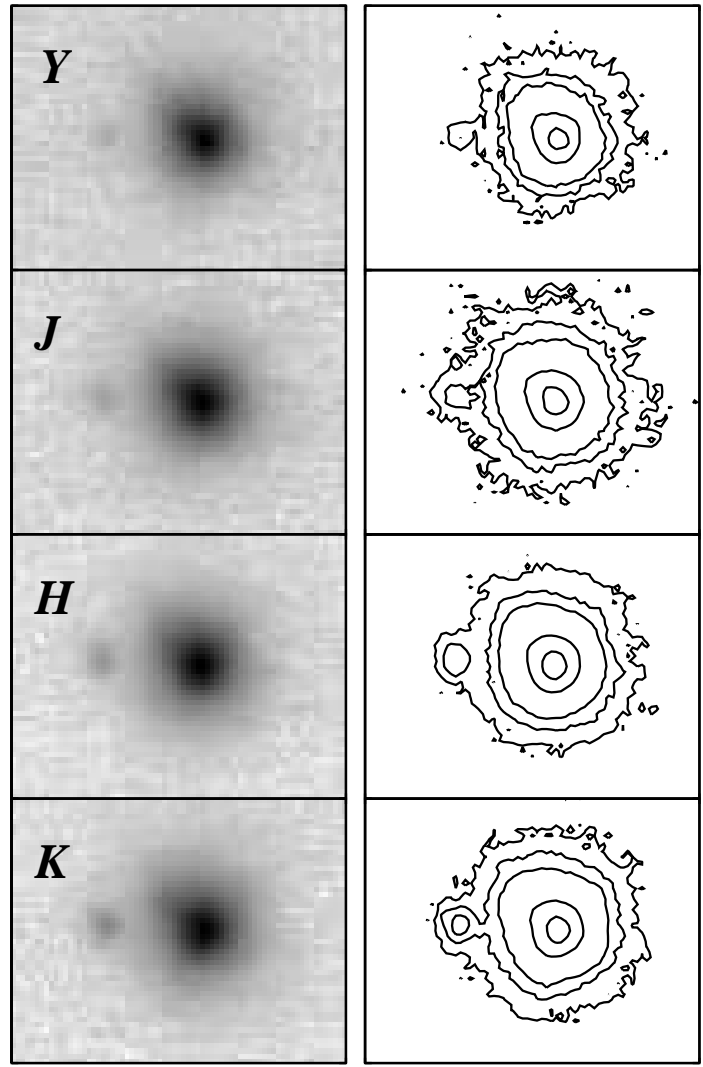}}
  \end{center}
  \caption{IRTF/SpeX guider camera $YJHK$ images of 1RXS~J2351+3127~AB.  The panels on the left show
  the stacked images of the system depicted with an asinh stretch.  The panels on the right
  show contours representing 50\%, 10\%, 1\%, 0.5\%, and 0.2\% of the peak flux from 
  the primary.\label{fig:guidedog} } 
\end{figure}

We obtained additional relative photometry of 1RXS~J2351+3127~AB with the guider camera 
on IRTF/SpeX on 2~Dec~2011~UT.  We imaged the system in the $Y$,\footnote{This filter is labeled ``$Z$'' in
the Guidedog GUI and instrument documentation, but with a bandpass of 0.95--1.11~$\mu$m it better resembles
the $Y$ filter described in \citet{Hillenbrand:2002p23339}.} $J$, $H$, and $K_S$ filters in nodded
patterns.  The details of our observations are listed in Table~\ref{tab:obs}.  Each frame was corrected for bad pixels
and cosmic rays, pair subtracted, and divided by a flat frame created from the science data
after masking out the system.  The images were then registered and stacked to extract relative 
photometry.
The companion is clearly visible in all filters and sits in the wing of the primary's PSF (Figure~\ref{fig:guidedog}).
We performed relative photometry by modeling each PSF as the  
sum of three elliptical Gaussians as described in \citet{Liu:2010p21195}.  
Uncertainties were computed by inserting and extracting artificial companions at same separation 
as the real object but different PAs.   The artificial companion was created by scaling the primary star 
to the same brightness as the real companion.
The resulting photometry is presented in Table~\ref{tab:photometry}, where quoted uncertainties
represent the standard deviation of several hundred realizations of this process.  Our IRTF $K$-band 
photometry disagrees with our Keck measurements by $\sim$4-$\sigma$.  Since the PSFs overlap 
in our IRTF data this discrepancy is probably a result of a slight 
systematic error or underestimated measurement uncertainty in our IRTF photometry.
We emphasize that while our IRTF data were obtained in excellent seeing conditions, our photometry and
spectroscopy of the companion may contain minor systematic errors as a result of its close proximity 
to the primary star.  Our Keck data taken with adaptive optics is expected to be much more reliable.

\subsection{Keck-II/OSIRIS $J$-Band Spectroscopy}{\label{sec:obsosiris}}

On 26~Dec~2011~UT we obtained Keck-II NGS-AO 1.18--1.35 $\mu$m 
spectroscopy of 1RXS~J2351+3127~B with the OH-Suppressing 
Infrared Imaging Spectrograph (OSIRIS; \citealt{Larkin:2006p5567}).
Unlike our IRTF/SpeX data, the combination of adaptive optics and an integral field unit enables resolved
spectroscopy of the companion without contamination from the primary star.  
We obtained three nodded pairs with exposures of 300~sec per position with the $Jbb$ filter and the 50~mas plate scale,
totaling 30~min of on-source data (Table~\ref{tab:obs}).
The resulting field of view was 0$\farcs$8$\times$3$\farcs$2 and the resolving power was $\sim$3800.
Immediately following our science observations we targeted the A0V star HD~78215 to correct for telluric features.

The data were reduced with the OSIRIS data reduction 
pipeline\footnote{Version 2.3: http://irlab.astro.ucla.edu/osiris/.} and the latest rectification matrix made available by the Keck Observatory.  Nodded pairs were used for
mutual sky subtraction. The spectra were extracted from the data cubes
using 3~pix (150~mas) circular apertures.  The individual spectra were first scaled to a median-combined 
spectrum, then deviant pixels were removed with a sigma-clipping algorithm at each wavelength using 
a 3~$\sigma$ threshold, and finally the spectra were combined by computing mean and standard errors
at each wavelength.  Telluric correction was performed with the \texttt{xtellcor\_general} routine 
in the Spextool spectroscopic reduction package.  At its native resolution the final spectrum
has a median S/N per pixel of $\sim$15.

\section{Results}{\label{sec:results}}

\subsection{Common Proper Motion}

\begin{figure}
  \vskip -1 cm
  \resizebox{\textwidth}{!}{\includegraphics{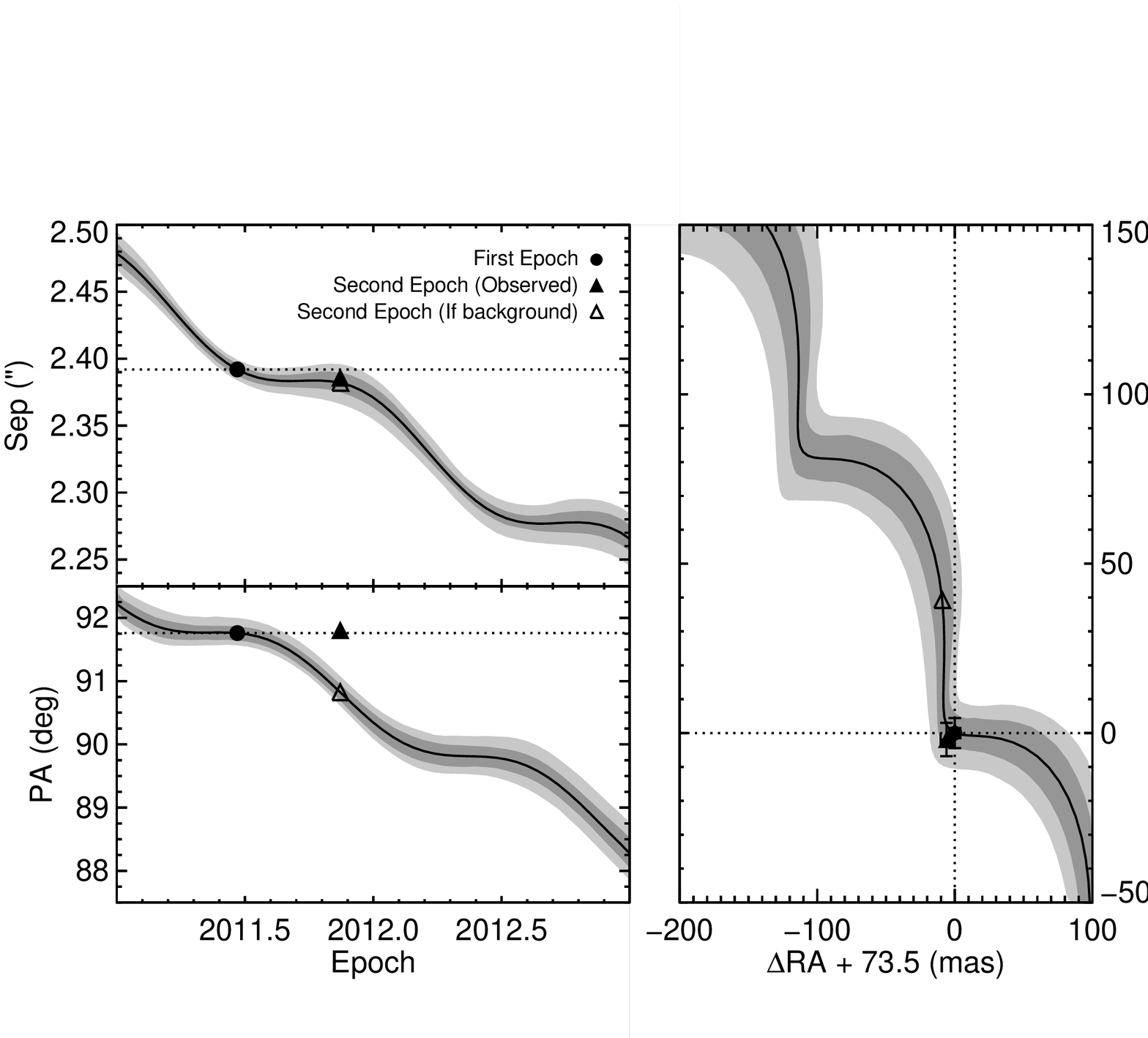}}
  \caption{Astrometry for 1RXS~J2351+3127~AB.  \emph{Left}: Separation (top) and 
  position angle (bottom) of the companion at two epochs in 2011.  The solid line shows the
  expected astrometry of a distant background object at the location of the companion in the first epoch data as a result
  of proper and parallax motion of the primary. 
  The gray shaded regions represent 1- and 2-$\sigma$ errors in the background tracks based
  on uncertainties in the proper motion, distance, and first epoch astrometry (solid circle).  The second
  epoch astrometry (solid triangle) is inconsistent with the expected position (in PA but not separation) if it were a 
  background object (open triangle).    The astrometric uncertainties are smaller than the 
  size of the symbols.
  \emph{Right}: Same as the left panel except for $\Delta$RA and $\Delta$Dec as seen on the sky
  ($\Delta$ refers to primary -- secondary position).  
  There is essentially no change in RA and Dec between the two epochs.\label{fig:background} } 
\end{figure}

We use the sky coordinates, distance estimate (50~$\pm$~10~pc, see Section~\ref{sec:dist}), and proper motion 
of 1RXS~J2351+3127~A along with first epoch astrometry of the candidate
companion to predict the
separation and position angle (or, equivalently, change in right ascension and declination)
of a distant background object  over time.  The results are shown in Figure~\ref{fig:background}, with shaded errors
at each epoch incorporating uncertainties in distance, proper motion, and first epoch astrometry.  
Our second epoch Keck astrometry is consistent with our first epoch measurements (within 3-$\sigma$)
and rules out the  background
hypothesis at 7-$\sigma$, proving the companion is comoving and very likely 
gravitationally bound to the primary.  The second epoch separation is indistinguishable from the expected
separation of a background object, but the position angle of a background object differs from the second
epoch measurement by 0.99~$\pm$0.13~deg.  Most of the uncertainty in this value is from the error in the
background model PA at that epoch rather than the measured PA of the companion.

\subsection{Distance}{\label{sec:dist}}

There are several distance estimates to the primary star 1RXS~J2351+3127~A in the literature.  \citet{Reid:2007p22752} 
use absolute magnitude-optical band strength index relations ($M_J$-TiO5 and $M_J$-CaH2) to arrive at an estimated
$M_J$ of 7.08~$\pm$~0.34~mag and a corresponding distance of 35.0~$\pm$~5.6~pc.   \citet{Riaz:2006p20030} use a slightly 
different $M_J$-TiO5 relation and arrive at a distance of 50~pc (we calculate an uncertainty of 20~pc using the quoted 
rms from their relation.)  Both estimates make use of field relations and assume the star is single.
Recently \citet{Schlieder:2012p23477} estimated a kinematic distance of 41.3~$\pm$3.2~pc for 1RXS~J2351+3127~A using the 
technique described in \citet{Lepine:2009p19553}, which assumes it is a member of the 
AB Dor YMG (see Section~\ref{sec:kinematics}).

The late-type companion can also be used to obtain an independent distance estimate for the system.
The most extensive compilation of field MLT objects with parallaxes was recently assembled by \citet{Dupuy:2012p23123}.
They measure an $M_{K_\mathrm{MKO}}$ value of 10.46~$\pm$~0.15 for L0 objects, which is the spectral type we 
adopt for 
1RXS~J2351+3127~B (Section~\ref{sec:spec}).  
Our $K'$ contrast measurement must first be converted to $K_\mathrm{MKO}$ to use this relation.  
The synthetic $K_\mathrm{MKO}$-$K'$ and 
$K_S$-$K'$ colors from our SXD spectrum of 1RXS~J2351+3127~A are $<0.02$ mag, so we 
assume $K_\mathrm{MKO}$$\sim$$K'$$\sim$$K_S$ for the primary.
Our prism spectrum of the companion yields a $K_\mathrm{MKO}$-$K'$ color of --0.09.  
Taking this into account and using the 2MASS magnitude
of the primary gives a $K_\mathrm{MKO}$-band magnitude of 13.92~$\pm$~0.05~mag and a distance of 49~$\pm$~5~pc.

If the system is young then field relations will underestimate the objects' luminosities and distances.
Assuming instead an age comparable to the Pleiades ($\sim$120~Myr), we can make use of photometry of known
late-type Pleiades members to infer an empirical $SpT$-absolute magnitude relation for young objects.  
Isolating Pleiades members from \citet{Bihain:2006p20005} with spectral types between M9 and L1 and using a cluster 
distance of 133~pc (\citealt{Soderblom:2005p23093})
yields an average $M_{K_\mathrm{MKO}}$ value of 10.0~$\pm$0.6~mag and a slightly larger distance of 60~$\pm$~17~pc
for 1RXS~J2351+3127~B.  Altogether we adopt a distance of 50~$\pm$~10~pc, with a younger
age favoring a larger distance.

\subsection{Age}

\subsubsection{X-ray Activity}

Low-mass stars have long been known to exhibit age-rotation-activity relationships spanning their
pre-main sequence and main sequence lifetimes (\citealt{Skumanich:1972p23130}).  
They are born with high angular momenta and fast
rotation rates, which induce strong magnetic fields and result in active coronal
and chromospheric emission (e.g., \citealt{Feigelson:1999p23306}; \citealt{West:2008p19562}).  
Angular momentum loss through stellar winds slows 
rotation and diminishes 
magnetic field strengths over time (\citealt{Feigelson:2004p23305}; \citealt{Barnes:2010p22812}; \citealt{Reiners:2012p23321}).  
Observationally this manifests as
an evolution of rotation rates (e.g., \citealt{Irwin:2011p22865}) and X-ray activity (e.g., \citealt{Preibisch:2005p330}); 
both are calibrated to young coeval clusters
and old field stars but have considerable dispersion for a given age and stellar mass.  

1RXS~J2351+3127~A was first identified in a large spectroscopic survey by \citet{Riaz:2006p20030} 
to locate new nearby M dwarfs using a combination of 2MASS and the $ROSAT$ All Sky Survey catalogs.  They found a high
fractional X-ray luminosity of log~$L_X$/$L_\mathrm{Bol}$ = --3.02, which corresponds to 
the saturation limit for low-mass stars (e.g., \citealt{Delfosse:1998p23131}; \citealt{Pizzolato:2003p23304}; \citealt{Wright:2011p23132}).  
How does this compare to typical values for young clusters?
\citet{Preibisch:2005p330} present a comprehensive 
analysis of the evolution of X-ray activity for various stellar mass bins between the ages of 
$\sim$1~Myr and several Gyr.   Fractional X-ray luminosities decline over time, but less precipitously
for low-mass stars compared to solar-type stars.  The median values for 0.1--0.5~\Msun \ stars only decrease by $\sim$0.4 dex
from the 1-10~Myr clusters to Hyades ages ($\sim$625~Myr), whereas 0.9--1.2~\Msun \ stars vary by $\sim$1.7 dex 
over the same timeframe.  Both stellar mass bins show a subsequent drop of $\sim$1.3~dex from the Hyades to the field age.
The fractional X-ray luminosity for 1RXS~J2351+3127~A is
well above typical values of even the youngest clusters.  The cumulative distributions  
for low-mass stars (0.1-0.5~\Msun) from \citet{Preibisch:2005p330} indicate that  
even the most active tail of field objects never reaches values of --3.0, although a more precise age determination is
difficult using log~$L_X$/$L_\mathrm{Bol}$ alone since the distributions for young clusters overlap.

\begin{figure}
  \resizebox{\textwidth}{!}{\includegraphics{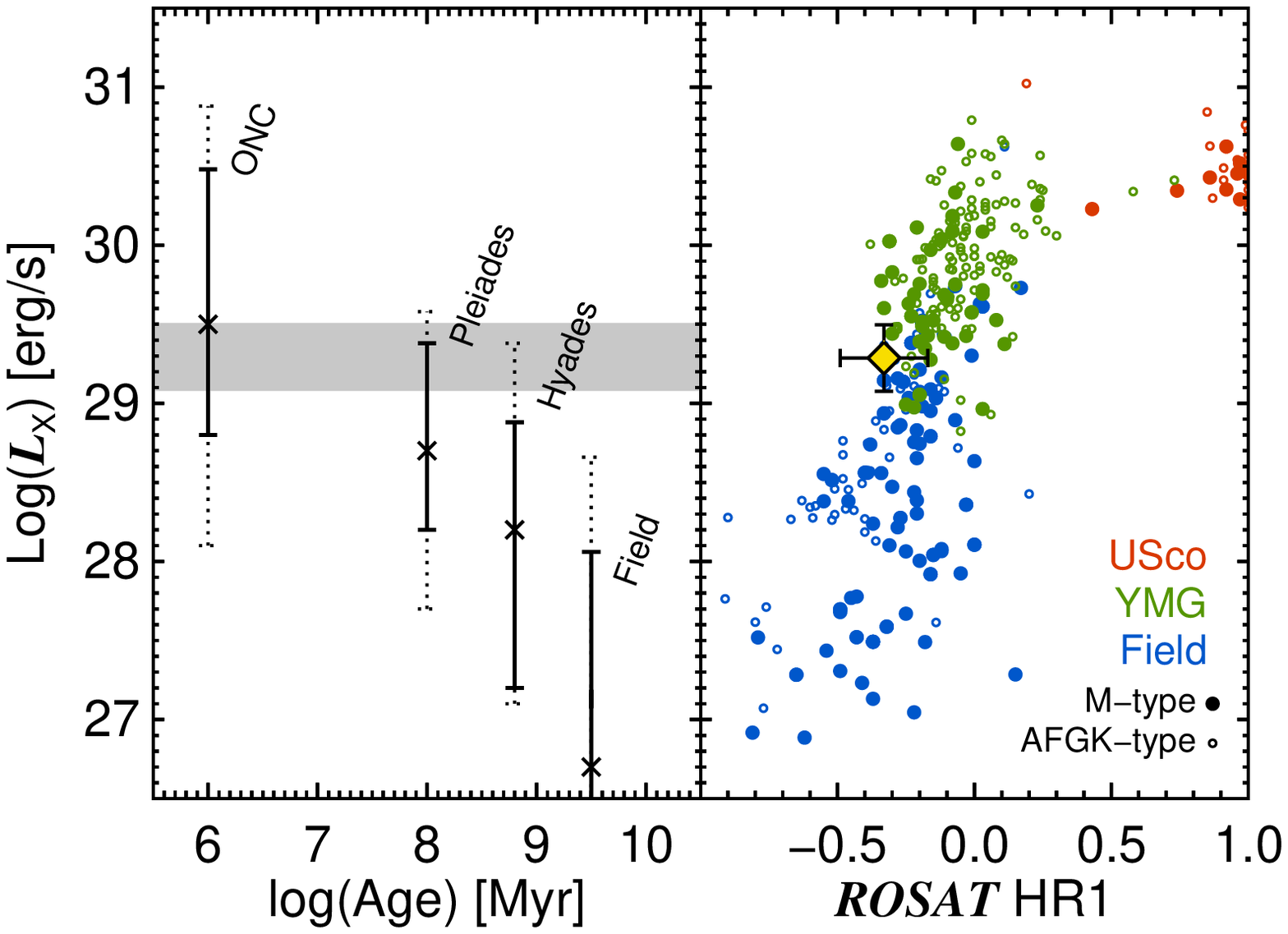}}
  \caption{  \emph{Left}: Distributions of X-ray luminosities for low-mas stars (0.1--0.5~\Msun) 
  as a function of age from \citet{Preibisch:2005p330}. 
  Solid and dashed lines represent 68.3\% and 95.4\% ranges about the median for each distribution, respectively.  
  Note that \citet{Preibisch:2005p330} applied a conversion factor to compare the X-ray luminosities from $ROSAT$ for the 
  Pleiades, Hyades, and field populations to the results for the ONC from $Chandra$.  We converted these values back
  into $ROSAT$ bandpasses and did the same for the ONC data.  The X-ray luminosity of 1RXS~J2351+3127~A is higher than
  the low-mass field population and most Hyades members.  
  \emph{Right}: X-ray luminosity vs. $ROSAT$ hardness ratio 1 for USco members (red;  \citealt{Preibisch:2008p22220}), YMG
  members (green;  \citealt{Torres:2008p20087}), and field stars (blue; \citealt{Schmitt:2004p23662}).   M-type stars are 
  plotted as filled circles and AFGK-type stars are shown 
  with smaller open circles.  Young stars have high X-ray luminosities and hardness ratios near zero,
  although we note that non-detections are not taken into account here.  1RXS~J2351+3127~A is consistent with $\sim$10-100~Myr
  YMG members assuming a photometric distance of 50~$\pm$10~pc.    \label{fig:xlum} } 
\end{figure}

The cumulative distributions of X-ray luminosities (log~$L_X$) from \citet{Preibisch:2005p330}  
are somewhat less degenerate than for fractional X-ray luminosities.  Using the count rate to flux conversion 
factor from \citet{Fleming:1995p23307} and assuming a distance of 50~$\pm$~10~pc yields log~$L_X$=29.3~$\pm$~0.2 erg/s
for 1RXS~J2351+3127~A.  In Figure~\ref{fig:xlum} (left panel) we compare this to the distribution of luminosities for 
various populations from \citet{Preibisch:2005p330}.
1RXS~J2351+3127~A is consistent with the ONC, Pleiades, and the most X-ray luminous members of the Hyades,
but is inconsistent with field objects.

\begin{figure}
  \vskip -1 cm
  \begin{center}
  \resizebox{5in}{!}{\includegraphics{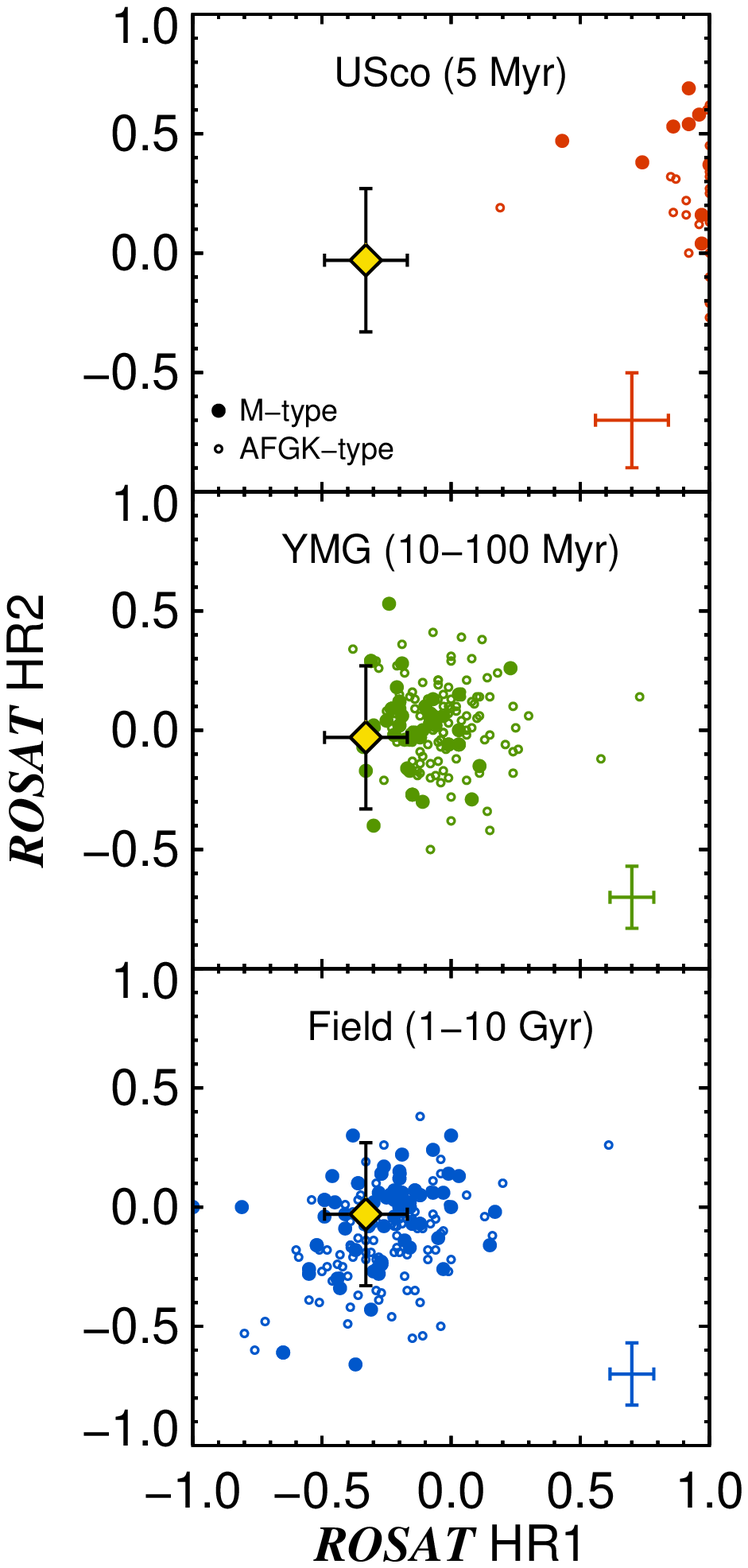}}
  \end{center}
  \caption{$ROSAT$ PSPC hardness ratios for members of Upper Scorpius 
  (top; \citealt{Preibisch:2008p22220}), YMG members (middle; \citealt{Torres:2008p20087}), and
  field stars within 15~pc (bottom; NEXXUS~2 catalog: \citealt{Schmitt:2004p23662}).  M-type stars are plotted as filled 
  circles and AFGK-type stars are shown with smaller open circles.  Hardness ratios soften over time 
  (\citealt{Kastner:2003p23108}) and can be used as a rough proxy for age. 1RXS~J2351+3127~A is consistent
  with YMG members and field objects.  Typical uncertainties are shown in the bottom right of each panel.  \label{fig:hr} } 
\end{figure}

The $ROSAT$ Position Sensitive Proportional Counter (PSPC) instrument had modest energy resolution, enabling a 
rough measurement of the
X-ray spectrum using hardness ratios as defined in \citet{Voges:1999p22945}.  
Stellar X-ray emission softens with age as a result of decreasing absorption by foreground gas at  
very early ages ($\lesssim$10~Myr; \citealt{Neuhauser:1995p22909}) and an evolving coronal gas temperature 
at intermediate and old ages ($\sim$10~Myr--10~Gyr; \citealt{Kastner:2003p23108}), so hardness ratios can be used
as a crude indicator of age. 
For a comparison sample we searched the RASS Bright Source Catalog (\citealt{Voges:1999p22945}) for 
X-ray counterparts within 40$''$ of stars in two age
bins:  Upper Scorpius members ($\sim$5~Myr) from \citet{Preibisch:2008p22220}; and AB~Dor, $\beta$~Pic, Columba, Tuc-Hor, 
and  TWA YMG members ($\sim$10-100~Myr) from \citet{Torres:2008p20087}.  In addition, we use 
the NEXXUS~2 catalog
(updated from \citealt{Schmitt:2004p23662}; C.~Liefke, private communication) for an older sample 
($\sim$1-10~Gyr) of field stars with RASS detections, 
limiting distances to $<$15~pc.
Figure~\ref{fig:hr} shows the resulting HR1/HR2 distributions;
USco members have very hard HR1 values near 1.0, the intermediate age sample has HR1 values 
between $\sim$--0.4 and +0.3, and field M dwarfs have values between $\sim$--0.5 and 0.0.
The hardness ratio of 1RXS~J2351+3127~A (HR1=-0.33~$\pm$0.16) is consistent with both YMG members
and M dwarfs in the field.

We also compute X-ray luminosities for the USco, YMG, and field populations 
and show the log~$L_X$ vs. HR1 distributions 
in Figure~\ref{fig:xlum} (right panel).  Note that not all of queried objects had X-ray counterparts, and since the RASS is a 
flux-limited survey the samples and distributions for each age group represent the most X-ray luminous members
of each population.
Deeper pointed observations of young compact clusters show a much larger spread in X-ray luminosities at a given
age, spanning, for example,
almost 3 orders of magnitude for the $\sim$1~Myr ONC cluster (\citealt{Preibisch:2005p330}).
(These results also suggest that all-sky searches which require X-ray detections to identify nearby young stars are missing a 
large fraction of young X-ray-faint members.)
The X-ray luminosity of 1RXS~J2351+3127~A appears to be inconsistent with field M dwarfs, agreeing better with young moving
group members which have ages of $\sim$10-100~Myr.

\citet{Shkolnik:2009p19565} and \citet{Schlieder:2012p23477} compute 
fractional X-ray fluxes for their samples of X-ray selected targets and derive values of
log~$F_X/F_J$ = --2.23 and log~$F_X/F_{K_S}$=--2.06, respectively, for 1RXS~J2351+3127~A.
Both values lie near the saturation
limit for early M dwarfs and are comparable to YMG members with ages
less than the Pleiades ($\sim$120~Myr).  \citet{Shkolnik:2009p19565} obtained two epochs of high resolution 
spectroscopy of 1RXS~J2351+3127~A and rule out spectroscopic binarity as 
the source of increased activity.  They also found a CaH band strength suggesting low surface
gravity but did not detect Li, leading to a a likely age of 20-150~Myr.

\subsubsection{UV Activity}

Chromospheric activity produces a wealth of emission lines and continuum flux at UV wavelengths
(e.g., \citealt{Robinson:2005p23314}; \citealt{Pagano:2009p23311}).  This activity decays over time 
(\citealt{Simon:1985p23319}; \citealt{Ribas:2005p23320}; \citealt{Findeisen:2011p22756}),
making excess UV emission a good tracer of magnetic field strength and age.  The \emph{Galaxy Evolution Explorer} 
($GALEX$; \citealt{Martin:2005p23310}) space telescope mapped most of the sky in near-UV (NUV) and far-UV (FUV) bands
and the resulting catalog (\citealt{Morrissey:2007p22251}) represents a rich resource to identify nearby young stars 
that exhibit UV excesses (\citealt{Shkolnik:2011p21923}; \citealt{Rodriguez:2011p21813}; 
\citealt{Schlieder:2012p23477}).

1RXS~J2351+3127~A was detected in both $NUV$ and $FUV$ bands of $GALEX$ (Table~\ref{tab:photometry}).  Here we compare
its UV emission to several empirically calibrated UV/near-infrared colors from the literature.
1RXS~J2351+3127~A exhibits high fractional UV fluxes relative to its $J$-band and $K_S$-band fluxes with values consistent
with local association members (\citealt{Shkolnik:2011p21923}; \citealt{Schlieder:2012p23477}).  \citet{Rodriguez:2011p21813}
find distinct loci for field and YMG populations in both $NUV$--$V$ vs. $V$--$K$ and $NUV$--$J$ vs. $J$--$K$ planes.  
Lacking a reliable $V$-band magnitude for 1RXS~J2351+3127~A, we can use the typical $V$--$K$ color for an M2V 
dwarf of 4.11~mag (\citealt{Tokunaga:2000p21771}) to provide an estimate.  Combining this with the 
2MASS $K_S$-band magnitude of 1RXS~J2351+3127~A yields a value of $V$=13.1~mag.  Comparing its 
$NUV$--$V$ color of $\sim$6.9~mag
to Figure~2 of Rodriguez et al. shows that 1RXS~J2351+3127~A is clearly discrepant from the field population and
again sits along the locus of YMG objects.  Likewise, its $NUV$--$J$ color (10.15~mag) is $\sim$1.5 magnitudes bluer than
the field population for its $J$--$K_S$ color (0.85~mag) based on Figure~4 of Rodriguez et al.  It 
is also bluer than Hyades sequence
based on Figure~7 of \citet{Findeisen:2011p22756}.  Although the scatter is quite large, \citet{Findeisen:2011p22756}
derive an empirical calibration for log(age) vs. $FUV$--$J$ and $J$--$K$ colors (their Equation~10), which yields
an age of 36$^{+53}_{-21}$~Myr.  Altogether, the UV fluxes of 1RXS~J2351+3127~A point to an age that is confidently less
than the Hyades (625~Myr) and likely between 10-150~Myr.

\subsubsection{Kinematics}\label{sec:kinematics}

Both \citet{Schlieder:2012p23477} and Shkolnik et al. (2012, submitted) independently find that 1RXS~J2351+3127~A
is a likely member of the AB~Dor YMG.  Schlieder~et~al. use the method 
of \citet{Lepine:2009p19553} to compute a kinematic distance and predict a radial velocity
assuming cluster membership.  Their kinematic distance for 1RXS~J2351+3127~A (41.3~$\pm$~3.2~pc) matches 
our photometric distance of the system (50~$\pm$~10~pc), and their predicted radial velocity (--14.0~$\pm$~1.3~km/s)
is in excellent agreement with the measured value of --13.5~$\pm$~0.6~km/s by Shkolnik et al.  
  The radial velocity of 1RXS~J2351+3127~A was measured from two high
resolution spectra ($\lambda$/$\Delta \lambda$ $\sim$ 68,000) obtained by  \citet{Shkolnik:2009p19565} 
on 14~Aug~2006~UT and 5~Oct~2006~UT with 
the \'{E}chelle SpectroPolarimetric Device for the Observation of Stars at the Canada-France-Hawaii Telescope. 
Details about the measurements will appear in 
Shkolnik et al. (2012, submitted).  In brief, the spectra were cross-correlated with an RV standard
with a similar spectral type and the radial velocity was derived from Gaussian fits to the cross-correlation functions.

$UVWXYZ$ values for 1RXS~J2351+3127~A with respect to the Sun's space motion and position are listed in Table~\ref{tab:properties}.   
Given the large range of photometric distances for the system, we also compute space motions for distances of 35, 45, 55, and 65~pc, which are
plotted in Figure~\ref{fig:uvw} relative to YMGs from \citet{Torres:2008p20087}.  For distances between $\sim$35 and 50~pc
the kinematics of 1RXS~J2351+3127~A agree well with the AB Dor group and sit in the cluster center at $\sim$45~pc.
Distances larger than 55~pc are discrepant with the cluster.   Although membership is highly probable, a parallax for the system and verification using cluster convergence-point methods (e.g., \citealt{Torres:2006p19650}; \citealt{Mamajek:2005p19635}; \citealt{Galli:2012p23323}) 
is essential for unambiguous association.

\subsubsection{Age Summary}\label{sec:agesummary}

The high X-ray and UV emission from 1RXS~J2351+3127~A point to an age significantly younger than the Hyades.
A more precise age determination from high energy emission alone is hindered by the large intrinsic scatter from young
moving group members.  The $UVW$ kinematics of 1RXS~J2351+3127~AB based on its photometric distance
are consistent with the AB~Dor moving group,
which has an age comparable to the Pleiades.  However, a parallax is required to verify cluster membership.  
Very young ages ($\lesssim$10~Myr) can be excluded based on the
morphology of the near-infrared spectrum of 1RXS~J2351+3127~B (Section~\ref{sec:spec}) and the lack
of Li in the primary. 
Altogether we adopt two age estimates for 1RXS~J2351+3127~AB: 50--150~Myr assuming AB~Dor membership, 
and a conservative estimate of 50--500~Myr
if the system does not belong to that cluster.

\subsection{Spectral Properties and Classification}{\label{sec:spec}}

\subsubsection{Low-Resolution Prism Spectrum}

Our 0.8-2.5~$\mu$m SpeX prism spectrum of 1RXS~J2351+3127~B shows typical features of 
late~M- and early~L-type objects (i.e., deep 1.5 and 1.9~$\mu$m steam bands; strong 2.3~$\mu$m CO absorption; 
and (blended) \ion{Na}{1}, \ion{K}{1}, and FeH features in $J$-band; \citealt{Cushing:2005p288})
despite clear and significant contamination from the primary
star at $\lambda \lesssim$1.2~$\mu$m (see Section~\ref{sec:prismobs}).  Here we attempt to use the uncontaminated 
(or minimally contaminated) regions of the spectrum for the purposes of
spectral classification.  Our comparison spectra come from the SpeX Prism Spectral Library.\footnote{Maintained by Adam Burgasser at http://pono.ucsd.edu/$\sim$adam/browndwarfs/spexprism.}  We tried fitting several spectral regions of 1RXS~J2351+3127~B to 
a sample of 618 published M, L, and T dwarf spectra which were also obtained with IRTF/SpeX in prism mode.  The $\chi^2$ 
statistic was used as a goodness-of-fit metric, and we ignored measurement uncertainties in the library spectra.

The best-fitting object to the entire 1.20--2.45~$\mu$m region is the intermediate-age
M9.0 (optical spectral type) brown dwarf LP~944-20 (Figure~\ref{fig:prism}).
Optical spectroscopy of LP~944-20 by \citet{Tinney:1998p23326} revealed Li absorption, indicating it is
a young substellar object (see also \citealt{Pavlenko:2007p23324}).  Follow-up studies uncovered evidence 
for X-ray flaring (\citealt{Rutledge:2000p23328}), 
quiescent radio emission (\citealt{Berger:2001p23329}), possible photometric variability (\citealt{Tinney:1999p23327}), 
and optical, but not infrared, periodic radial velocity variations (\citealt{Martin:2006p23330}).  All of this indicates 
that LP~944-20 harbors an unusually strong magnetic field for its spectral type.  
\citet{Ribas:2003p23325} found that LP~944-20 is a likely member of the Castor YMG (which includes
the well-studied stars Vega and Fomalhaut) based on its space motion.  Age estimates for the Castor YMG
range from $\sim$200--400~Myr (\citealt{Barrado:1998p23331}; \citealt{Torres:2002p23332}; 
\citealt{Ribas:2003p23325}), i.e., intermediate in age between the Pleiades (125~Myr) and the Hyades~(625~Myr).

The near-infrared prism spectrum of LP~944-20 itself (\citealt{Burgasser:2008p14471}) reveals only subtle hints of youth.  
One of the most
discriminating features exhibited by young brown dwarfs at low spectral resolution is a triangular-shaped 
$H$-band (e.g., \citealt{Lucas:2001p22099}; \citealt{Allers:2007p66}).  This feature is readily seen in Figure~\ref{fig:prism}, which shows the 
spectrum of LP~944-20 compared to the L0 optical standard 2MASS~J0345432+254023 (\citealt{Burgasser:2006p23333})
and the young M9.5 TWA brown dwarf 2MASS~J11395113--3159214 (\citealt{Looper:2007p3713}).
LP~944-20 has an intermediate $H$-band shape reflecting its adolescent age.  This subtle spectral peculiarity is 
shared by 1RXS~J2351+3127~B and suggests a comparable age to LP~944-20 based on this morphology alone.
(The 1.4--1.8~$\mu$m region of 1RXS~J2351+3127~B, which samples the H$_2$O band depth
and the entire $H$-band region, is also best-fit by the spectrum of LP~944-20.)
However, since the spectral properties of brown dwarfs at intermediate ages are not well-calibrated, it is unclear
how quickly these features evolve and therefore what robust age constraints can be obtained from this 
method.  

Separate fits to individual bands produce best-fit spectral types of M9--L1.5. 
The 1.15--1.34~$\mu$m region yields the L0 (optical type: \citealt{West:2008p19562})/M9 
(NIR type: \citealt{Kirkpatrick:2010p21127}) object 2MASSJ~12490872+4157286.  The best match 
to the 1.50--1.80~$\mu$m region is the L1.5 (NIR type: \citealt{Kirkpatrick:2010p21127}) 
object 2MASS~J01472702+4731142.  
The 2.0--2.45~$\mu$m region ($\sim$$K$) is best fit by the L0 (optical type: \citealt{Wilson:2001p18725}) object 
HD~89744~B from \citet{Burgasser:2008p14471}.\footnote{The primary star HD~89744~A (F7IV/V) was classified 
as a likely AB~Dor member by 
\citet{LopezSantiago:2006p18285} based on its space motion. However, it is also an exoplanet-host star and
independent age estimates point to an older age of 1--3~Gyr, arguing against AB~Dor membership 
 (see Table 3, footnote i of \citealt{Evans:2011p23125} for a summary).
The X-ray luminosity based on the count rate and hardness
ratio (HR1=--0.58) listed in the Second $ROSAT$ PSPC Catalog is a modest 28.0~erg/s, which is comparable to field stars.}  
Finally, we note that the gravity-independent classification index from \citet{Allers:2007p66} yields
a spectral type of M8.4$^{+1.1}_{-1.0}$, which is consistent with our previous estimates.

\subsubsection{Moderate-Resolution SXD and OSIRIS Spectra}

In Figure~\ref{fig:sxd} we compare our SXD spectrum to late-M and early-L dwarfs from the 
IRTF Spectral Library (\citealt{Cushing:2005p288}).  1RXS~J2351+3127~B most closely resembles
the L0.5 and L1 field objects.  A comparison of individual bands to the templates likewise
suggests M9--L1.  Like our prism spectrum, there is some contamination from the primary
at shorter wavelengths, which results in a slightly bluer spectrum and artificially diminished
line strengths.  The $H$ and $K$ bands were less contaminated by the primary  
in the raw SXD data.

On the other hand, our OSIRIS spectrum was obtained with AO and the companion was well separated from the primary.
The $J$-band spectrum is shown in Figure~\ref{fig:osiris} relative to M8--L3 field templates.  The best overall match
is the L2 template, although L1--L3 objects are also good fits.  The depth of the 1.25~$\mu$m \ion{K}{1} gravity-sensitive 
lines (e.g., \citealt{McGovern:2004p21811}; \citealt{Kirkpatrick:2006p20500}; \citealt{Allers:2009p18424}) 
do not appear particularly shallow relative to the field objects, which excludes very young ages ($\lesssim$10~Myr).
Recently \citet{Geiler:2012p23335} presented NIR spectroscopy of a new M8~$\pm$~1 companion to the
Pleiades member \ion{H}{2}~1348.  Their OSIRIS $J$-band spectrum also lacks shallow alkali lines, which
suggests that any impact on these lines from low surface gravity occurs at younger ages, at least at this temperature.
Altogether we adopt a spectral type of L0$^{+2}_{-1}$ for 1RXS~J2351+3127~B.  The asymmetric error bars
reflect the slightly earlier type suggested by the prism and SXD spectra compared to the OSIRIS spectrum.

\subsection{Physical Properties}

The mass of 1RXS~J2351+3127~B can be estimated from its luminosity and age using substellar cooling models.
Assuming a distance of 50~$\pm$~10~pc, the $K$-band bolometric correction from 
\citet{Golimowski:2004p15703} yields a luminosity of log~$L/L_{\odot}$~=~--3.6~$\pm$~0.2 
(the uncertainty incorporates intrinsic scatter in the relation,
photometric errors, and a spectral type accuracy of one subtype).\footnote{Distances of 
\{35, 45, 55, 65\}~$\pm$~10~pc give log~$L/L_{\odot}$~=\{--3.9$\pm$0.3, --3.6$\pm$0.2 ,--3.5$\pm$0.2, --3.3$\pm$0.1\}.}
The age estimate of the system depends on whether it is an AB~Dor member, in which case the age of
$\sim$50--150~Myr (see Section~\ref{sec:discussion}) can be
leveraged from the entire cluster.  
If it is not a member then the constraints are much poorer, but probably lie between 50--500~Myr based on the
high energy emission from the primary.
We arrive at masses of 32~$\pm$~6~\Mjup \ and 50~$\pm$~11~\Mjup \ for age ranges of 50--150~Myr and 50--500~Myr, 
respectively, based on a grid of finely interpolated evolutionary models from 
\citet{Burrows:1997p2706}.\footnote{Quoted masses
represent the medians and standard deviations of the resulting mass distributions assuming uniform input 
distributions for the luminosity and ages.  
Normally-distributed input values of --3.6~$\pm$~0.2~dex and 100~$\pm$~25~Myr (300~$\pm$150~Myr) give a similar 
mass of 35~$\pm$~7~\Mjup \ (56~$\pm$~11~\Mjup).  We also compute masses using interpolated Lyon models from 
\citet[Dusty]{Chabrier:2000p161} and \citet[Cond]{Baraffe:2003p587}, as well as 
\citet[both clear and cloudy versions]{Saumon:2008p14070}.  The resulting mass estimates are all consistent
within a few \Mjup, indicating that the dominant sources of error are the uncertainty in the age and distance 
rather than the choice of models.}
Figure~\ref{fig:lumage} shows the influence of the distance estimate on the inferred mass of 1RXS~J2351+3127~B,
which lies between ~$\sim$25--40~\Mjup \ for distances of 35--65~pc assuming an age of 50--150~Myr.

The mass of the primary can be obtained from stellar evolutionary models.
1RXS~J2351+3127~A has a spectral type of M2.0~$\pm$~0.5 which corresponds to an effective temperature
of $\sim$3520~K (\citealt{Drilling:2000p21754}).  At 100~Myr (500~Myr), the models of \citet{Baraffe:1998p160}
give a mass of 0.45~\Msun \ (0.40~\Msun) for this temperature.  Since the mass is only weakly dependent on age
we adopt a value of 0.45~$\pm$~0.05~\Msun \ for 1RXS~J2351+3127~A.  Note that we have 
made use of solar metallicity evolutionary models here (with $Y$=0.275 and $L_\mathrm{mix}$=$H_P$); a non-solar composition and any
intrinsic errors in the evolutionary models will result in systematic errors in the inferred mass.
We also compute the luminosity
of the primary star using the $H$-band bolometric correction from \citet{Casagrande:2008p23483} and 
the system's photometric distance, which yields  log~$L/L_{\odot}$~=~--1.37~$\pm$~0.19.

\section{Discussion and Conclusions}{\label{sec:discussion}}
 
 The AB~Dor YMG was first recognized by \citet{Zuckerman:2004p22744} as a sparse but relatively nearby group of 
 young stars in the Local Association with common space motions.  
 Since its discovery there have been many attempts to identify additional members 
 (\citealt{LopezSantiago:2006p18285}; \citealt{Torres:2008p20087}; \citealt{Zuckerman:2011p22621}; 
 \citealt{McCarthy:2012p23334}), with 
 recent attention mostly focused on filling in the lower main sequence (\citealt{Shkolnik:2009p19565}; 
 \citealt{Schlieder:2010p21925}, 2012; Shkolnik et al. 2012).  
 Age estimates for the cluster vary considerably.  Studies of the AB~Dor quadruple system itself
 have yielded values of $\sim$30--100~Myr using theoretical isochrones 
 (\citealt{Close:2007p19548}; \citealt{Janson:2007p22748}; \citealt{Boccaletti:2008p22749}; \citealt{Guirado:2011p22745}).  
 However, comparisons of AB~Dor members to the Pleiades ($\sim$125~Myr)
and IC~2391 (35--50~Myr) clusters in color-magnitude diagrams by 
\citet[see also \citealt{Luhman:2006p1092}]{Luhman:2005p22437} 
make it clear that the AB~Dor group is older than IC~2391 and 
approximately coeval with the Pleiades.  Kinematic analysis of the cluster by \citet{Luhman:2005p22437}
and \citet{Ortega:2007p22746} support a common origin and age with the Pleiades.
In Figure~\ref{fig:abdor} we compare the color-magnitude sequence of AB~Dor members 
to Pleiades stars and brown dwarfs (see caption for details).  The sequences
line up well from the highest-mass B-type members down to the latest L dwarfs in both clusters,
supporting the older Pleiades-like age of AB~Dor.

\begin{figure}
  \resizebox{\textwidth}{!}{\includegraphics{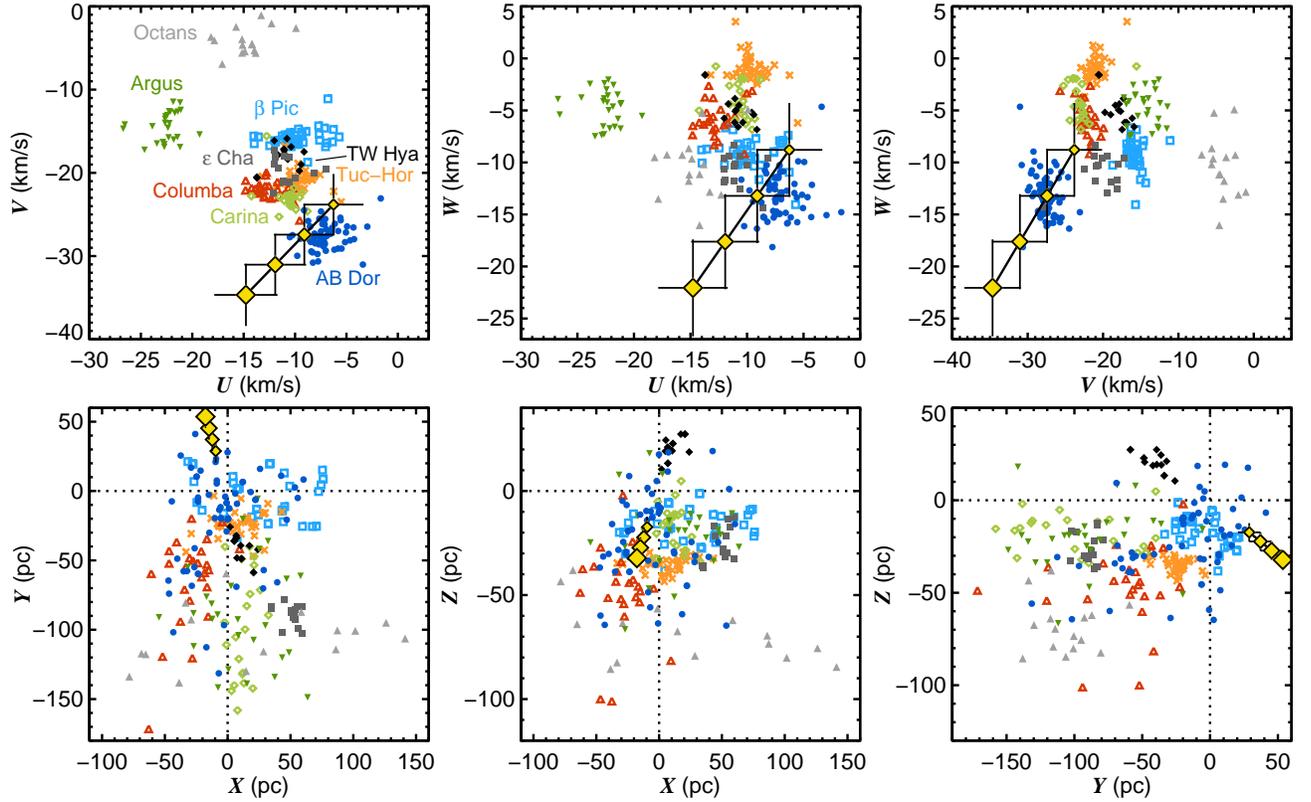}}
  \caption{Space motion and relative positions of nearby YMGs from  \citet{Torres:2008p20087}.  
  Distances (both measured parallaxes and kinematic estimates) are from Torres et al., while radial velocities
  are compiled from Simbad.  The Octans (filled triangles), Argus (filled down-facing triangles), $\beta$~Pic (open squares), 
  $\epsilon$~Cha (filled squares), TW~Hya (filled diamonds), Tuc-Hor (crosses), Columba (open triangles), 
  Carina (open diamonds), and AB~Dor (filled circles) groups occupy unique loci in $UVW$ space.  
  1RXS~J2351+3127~AB  is overplotted as a series of increasingly-larger yellow diamonds which represent distances
  of 35, 45, 55, and 65~pc.  The system's photometric distance of 50~pc coincides well with the AB~Dor YMG,
  but distances larger than $\sim$55~pc disagree with the kinematics of known members.  
  Error bars include uncertainties in the distance estimate (10~pc), proper motion, and radial velocity.
   The AB~Dor YMG is physically dispersed over a large region of sky; 1RXS~J2351+3127~AB is consistent with the cluster in $XYZ$ space,
   albeit near the border of where known members lie.  
  \label{fig:uvw} } 
\end{figure}

\begin{figure}
  \resizebox{\textwidth}{!}{\includegraphics{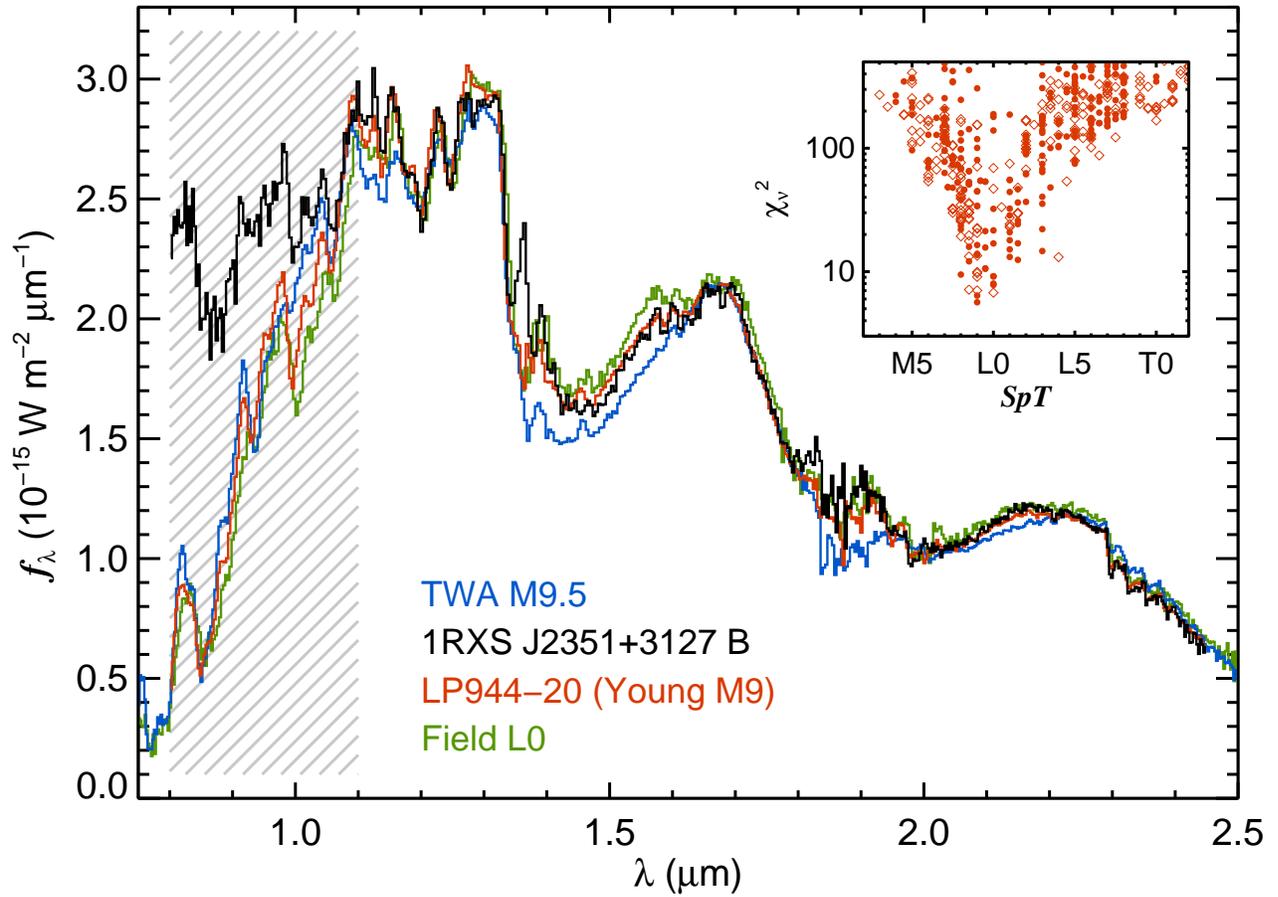}}
  \caption{IRTF/SpeX prism spectrum of 1RXS~J2351+3127~B (black).  The 1.15--2.45~$\mu$m region is best matched
  by the $\sim$200--400~Myr M9 brown dwarf LP~944-20 (red).  Contamination from the primary is evident at wavelengths
  shorter than $\sim$1.1~$\mu$m and is shown with gray shading.  The L0 optical standard 2MASS~J0345432+254023
  (green) and the young ($\sim$10~Myr) M9.5 TWA member 2MASS~J11395113--3159214 (blue) are shown for comparison.  The $H$-band
  region of 1RXS~J2351+3127~B and LP~944-20 appear intermediate between the old field object and the young brown dwarf,
  although the effect is subtle.  Note that LP~944-20 is scaled to 1RXS~J2351+3127~B by minimizing the $\chi^2$ value, while
  the field and young objects are normalized to the 1.68--1.70~$\mu$m region.  The inset shows the reduced $\chi^2$ values
  of fits to MLT dwarfs from the SpeX Prism Library plotted against spectral type.  When optical types (filled circles) are not available we use
  near-infrared types (open diamonds).   \label{fig:prism} } 
\end{figure}

\begin{figure}
  \resizebox{\textwidth}{!}{\includegraphics{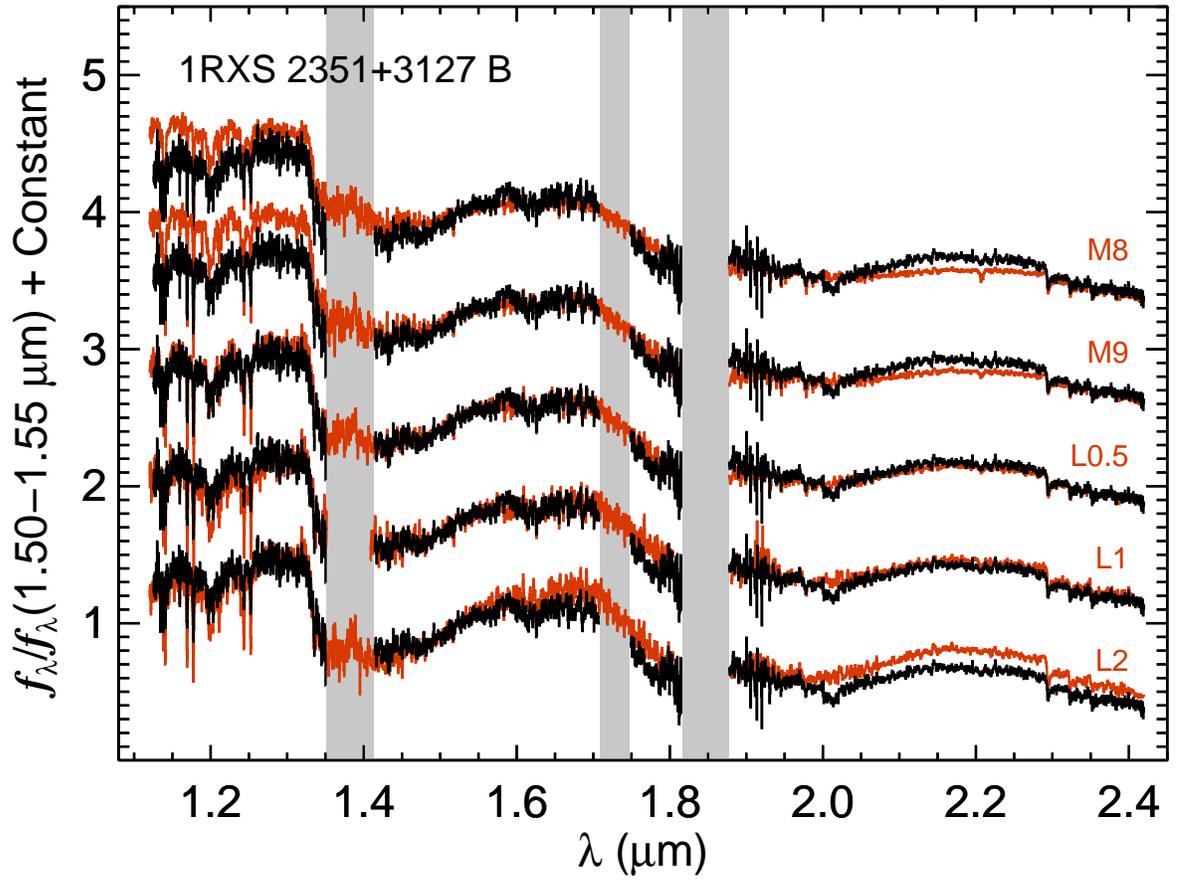}}
  \caption{IRTF/SpeX SXD spectrum of 1RXS~J2351+3127~B compared to field M and L dwarfs from 
  the IRTF Spectral Library (\citealt{Cushing:2005p288}).  The L0.5 and L1 templates are close matches to 
   1RXS~J2351+3127~B.  The spectra are normalized beetwen 1.50-1.55~$\mu$m and offset
   by a constant.  Gaps in the spectrum of  1RXS~J2351+3127~B are caused by strong telluric absorption 
   between 1.35--1.40~$\mu$m, a ghost feature between 1.71--1.75~$\mu$m caused by the placement on the 
   detector, and a gap between orders from 1.82--1.87~$\mu$m. \label{fig:sxd} } 
\end{figure}

\begin{figure}
  \begin{center}
  \resizebox{5in}{!}{\includegraphics{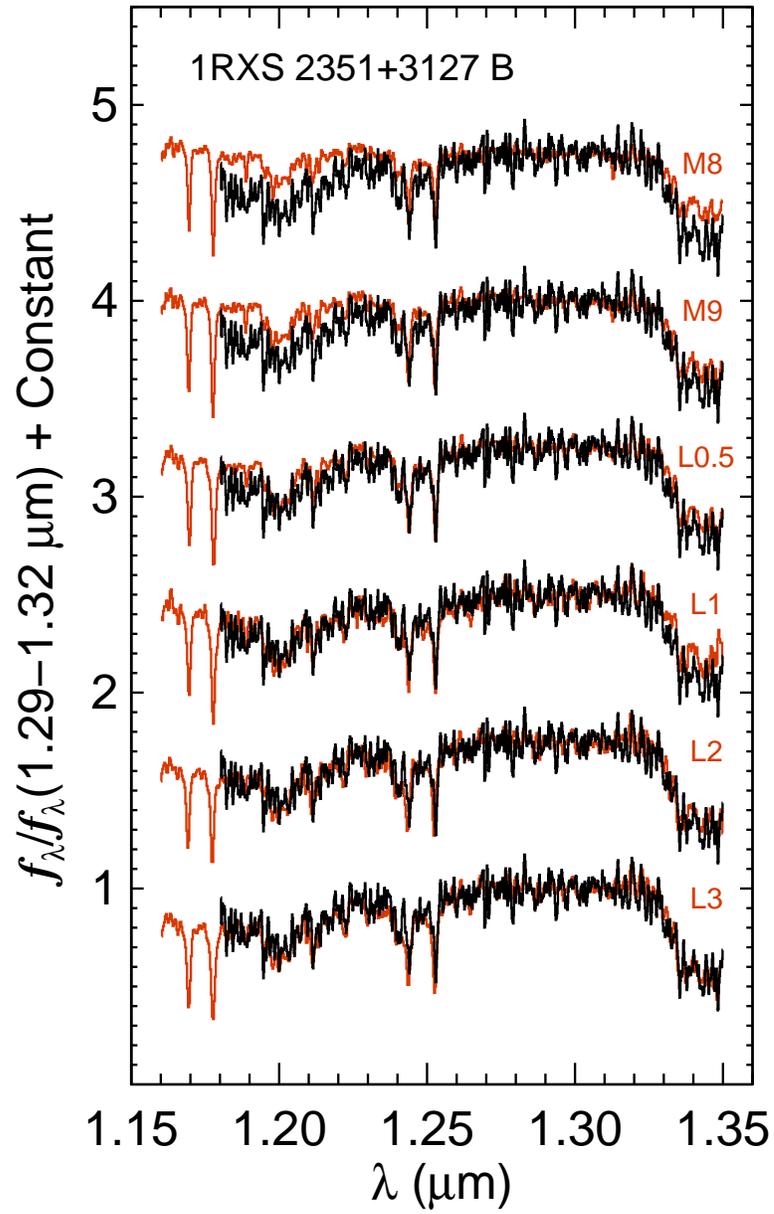}}
  \end{center}
  \caption{Keck-II/OSIRIS $J$-band spectrum of 1RXS~J2351+3127~B compared to field objects from the
  IRTF Spectral Library.  The depth of
  the 1.244/1.253~$\mu$m \ion{K}{1} lines are comparable to field objects, implying 1RXS~J2351+3127~B is not 
  exceptionally young ($\lesssim$10~Myr). The best matches are L1--L3 spectral types.\label{fig:osiris} } 
\end{figure}

\begin{figure}
  \resizebox{\textwidth}{!}{\includegraphics{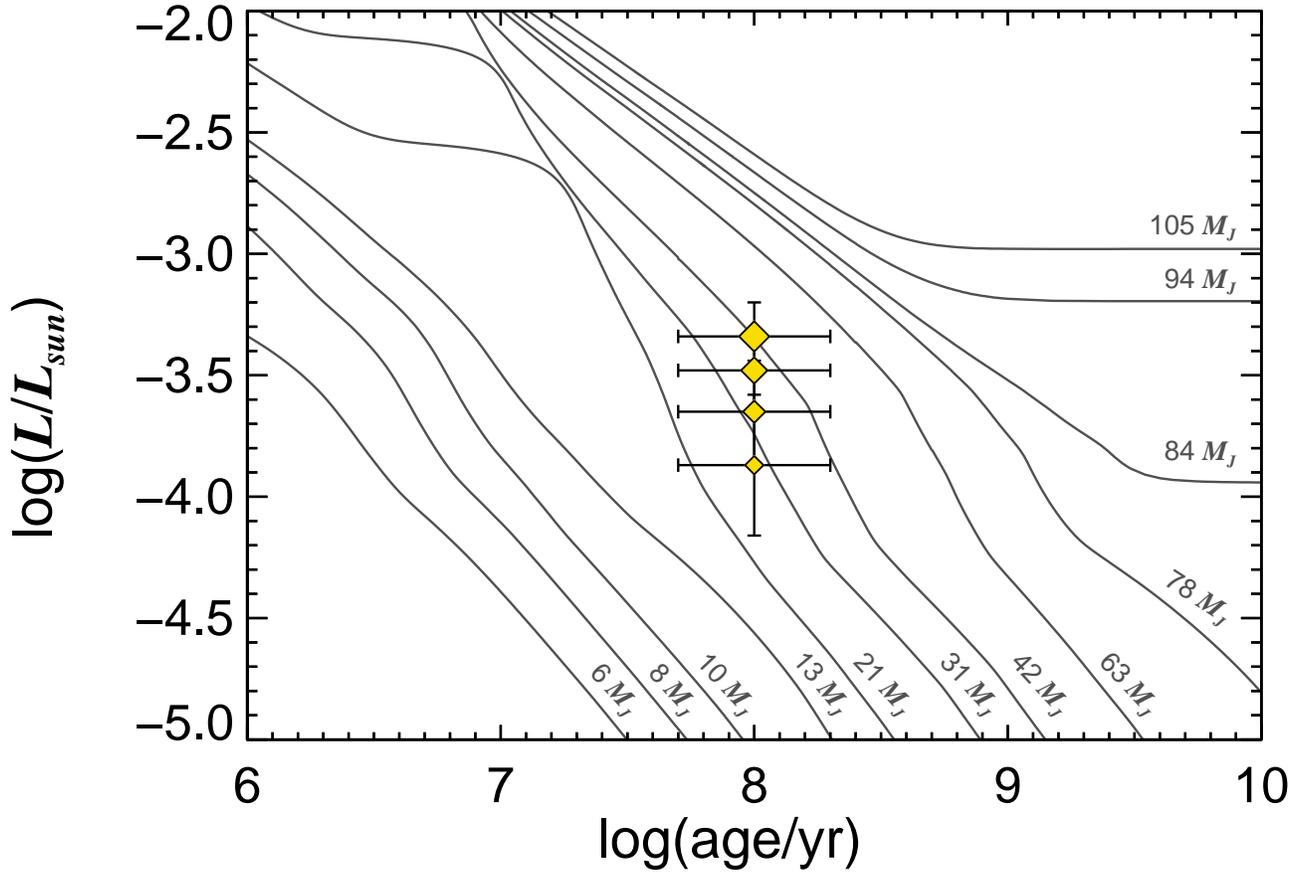}}
  \caption{The mass of 1RXS~J2351+3127~B based on the substellar evolutionary models 
  of \citet{Burrows:1997p2706}.  Yellow diamonds show the luminosity of 1RXS~J2351+3127~B for
  distances of 35, 45, 55, and 65~pc with increasing size indicating larger distance.  If 1RXS~J2351+3127~AB
  is a member of the AB~Dor moving group ($\sim$50--150~Myr), its photometric distance (50~$\pm$~10~pc)
  implies a mass of 32~$\pm$~6~\Mjup \ for the companion.  If the system is not a member and the
  age range is $\sim$50-500~Myr, the mass estimate increases to 50~$\pm$~11~\Mjup.  \label{fig:lumage} } 
\end{figure}

\begin{figure}
  \begin{center}
  \resizebox{4.5in}{!}{\includegraphics{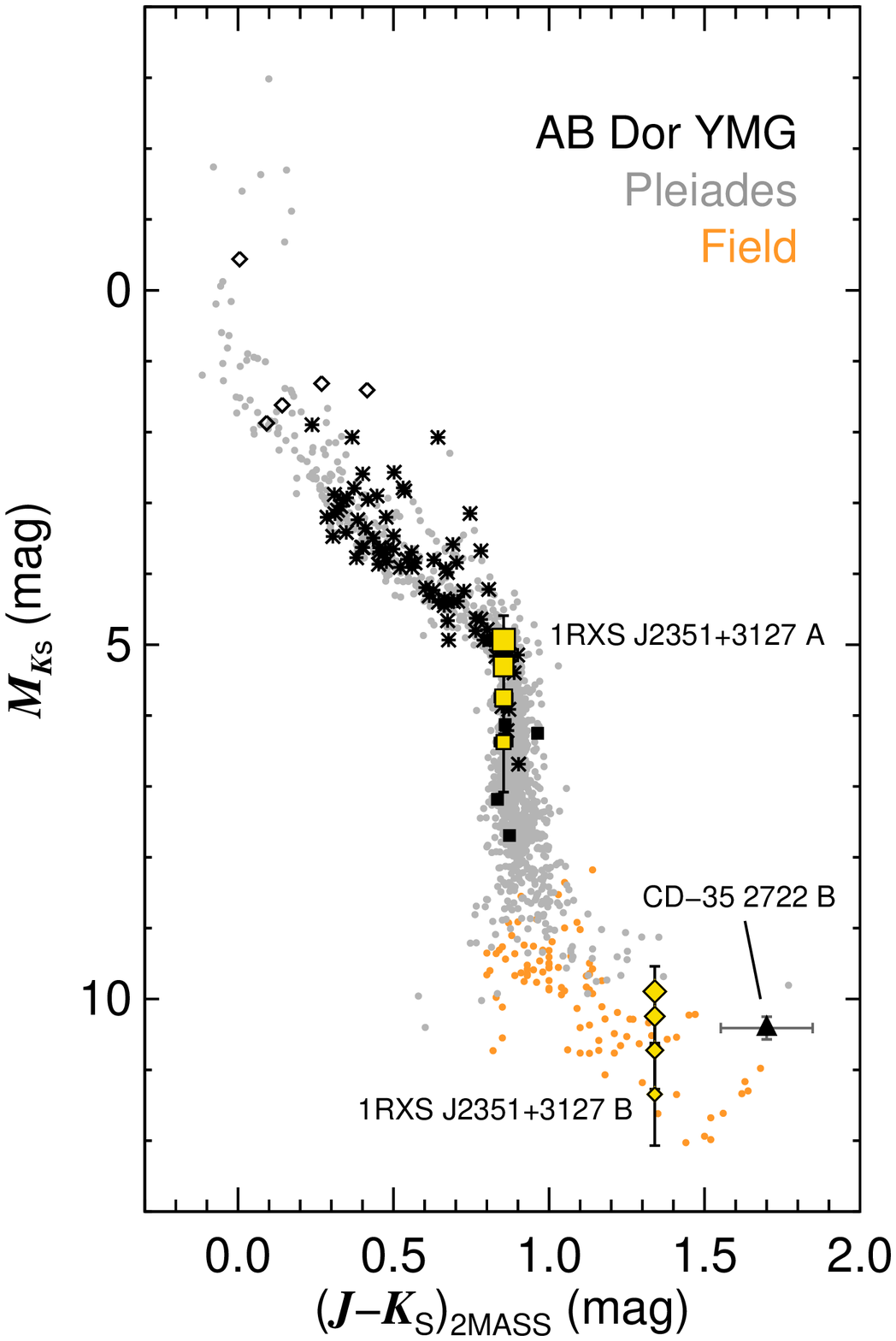}}
  \end{center}
  \caption{Color-magnitude diagram for the AB~Dor moving group.  The sequence of Pleiades members from 
  \citet{Stauffer:2007p19665} and \citet{Bihain:2010p21347} are shown in gray, and field ultracool 
  dwarfs from \citet{Dupuy:2012p23123} are plotted in orange.  The AB~Dor members are compiled from 
  \citet[stars]{Torres:2008p20087}, \citet[open diamonds]{Zuckerman:2011p22621}, and 
  Shkolnik et al. (2012, submitted; filled squares).  Except for the Torres et al. list, only objects 
  with parallaxes and good 2MASS photometry are plotted.  The AB~Dor sequence appears to be 
  indistinguishable from the Pleiades sequence, at least in $J$-$K$.  The positions of 1RXS~J2351+3127~AB
  are plotted with yellow squares (primary) and diamonds (companion) for distances of 35--65~pc.  
  The filled triangle shows CD--35~2722~B (\citealt{Wahhaj:2011p22103}, L4~$\pm$~1), which is the 
  latest-type member of AB~Dor currently known.  Its photometry was converted from the MKO to 2MASS systems
  using the relations from \citet{Leggett:2006p2674}.  Note that AB~Dor~C (M6~$\pm$~1) is not displayed because of the large
  uncertainties in its near-infrared colors (\citealt{Luhman:2006p1092}).  \label{fig:abdor}} 
\end{figure}

 A parallax for the 1RXS~J2351+3127~AB system is needed to confirm its association with AB~Dor.
  The large uncertainty in the photometric distance means the space motion of 1RXS~J2351+3127~AB is only marginally constrained.  
  At distances of 35--50~pc it agrees well 
 with the kinematics of the AB Dor group, but it is inconsistent at larger distances  
 given the relatively small internal velocity dispersion (a few km/s) of the moving group (Figure~\ref{fig:uvw}).
 If it does belong to the cluster then 1RXS~J2351+3127~B is the second lowest-mass member of AB~Dor after
 the L4 companion CD--35~2722~B (\citealt{Wahhaj:2011p22103}), which was recently discovered as part of the  
 Gemini NICI Planet-Finding Campaign (\citealt{Liu:2010p21647}).  
Note that 1RXS~J2351+3127~B should have a distance $\gtrsim$45~pc 
 for its position to be consistent with the later-type CD--35~2722~B in the color-magnitude diagram.
This could be a clue that 1RXS~J2351+3127~B is not in fact a member of AB~Dor since
distances larger than $\sim$50~pc produce space motions that differ from the cluster.  
As noted earlier, even if 1RXS~J2351+3127~AB are not members, independent
lines of evidence from the primary (high-energy emission) and the companion ($H$-band spectral morphology)
indicate the system is young ($\sim$100-500~Myr), and except for the oldest ages and the largest distances,
the companion is substellar.

 Already it is clear from the spectra of CD--35~2722~B obtained by  Wahhaj et al. that L dwarfs at the age 
of the AB~Dor moving group show few signs of youth.  Like 1RXS~J2351+3127~B, CD--35~2722~B has $J$-band
absorption features comparable to field objects, and the shape of the $H$ band is intermediate between
the youngest brown dwarfs and old field objects.  CD--35~2722~B has brighter absolute NIR magnitudes 
than field objects with
the same spectral type so a parallax for 1RXS~J2351+3127~B will also be useful to look for the same effect.
One of the few studies to examine the near-infrared spectral properties of Pleiades brown dwarfs was
carried out by \citet{Bihain:2010p21347}.  They obtained low-resolution spectra of M7--L3.5 Pleiades members
and most exhibited shapes similar to field objects, with a few suggesting somewhat more angular $H$-band
features with less 1.6~$\mu$m FeH absorption.  These trends appear to be consistent with CD--35~2722~B
and 1RXS~J2351+3127~B.
 
 A growing number of ultracool objects have been tied to the AB~Dor YMG.  In addition to 1RXS~J2351+3127~B (L0)
 and CD--35~2722~B (L4), which also orbits an early-M dwarf, 
 three isolated M7--M9 objects were recently found by \citet{Schlieder:2012p23477} 
 to be likely AB~Dor members, although parallaxes and radial velocities are needed for confirmation.  
 Including the low-mass M6 star AB~Dor~C (\citealt{Close:2005p23336}), a sequence of AB~Dor members
spanning a range of temperatures and masses below the hydrogen-burning limit is beginning to emerge.  
Since the spectral morphology
of brown dwarfs is sensitive to age, a detailed comparison of optical and near-infrared spectroscopy of AB~Dor
brown dwarfs to those in the Pleiades can eventually be used for relative age dating, perhaps 
with even greater precision than color-magnitude diagram comparisons.

  \acknowledgments

We thank the referee for helpful comments,
Joshua Schlieder for early access to his kinematic analysis of 1RXS~J2351+3127~A,
Katelyn Allers for helpful discussions about young brown dwarfs,
Eric Nielsen for the background track predictions, 
and Carolin Liefke for providing us with the NEXXUS~2 catalog.
It is a pleasure to thank the support astronomers and telescope
operators at Keck and IRTF who helped make this work possible:
Marc Kassis, Heather Hershley, Jim Lyke, Hien Tran, and John Rayner.
BPB and MCL have been supported by NASA grant NNX11AC31G and NSF grant AST09-09222.
ALK has been supported by NASA
through Hubble Fellowship grant 51257.01 awarded by STScI, which is
operated by AURA, INc., for NASA under contract NAS 5-26555.
We utilized data products from the Two Micron All Sky Survey, which is a joint project of the University of Massachusetts and the Infrared Processing and Analysis Center/California Institute of Technology, funded by the National Aeronautics and Space Administration and the National Science Foundation.
 NASA's Astrophysics Data System Bibliographic Services together with the VizieR catalogue access tool and SIMBAD database 
operated at CDS, Strasbourg, France, were invaluable resources for this work.
Finally, mahalo nui loa to the kama`\={a}ina of Hawai`i for their support of Keck and the Mauna Kea observatories.
We are grateful to conduct observations from this mountain.

\facility{{\it Facilities}: \facility{Keck:II (NIRC2, OSIRIS)}, \facility{IRTF (SpeX)}}

\newpage

\bibliographystyle{apj}

\clearpage

\clearpage
\newpage

\clearpage
\newpage

\clearpage
\newpage

\clearpage
\newpage

\clearpage
\newpage

\clearpage
\newpage

\clearpage
\newpage

\clearpage
\newpage

\clearpage
\newpage

\clearpage
\newpage

\newpage

\begin{deluxetable}{lccccc}
\tablewidth{0pt}
\tablecolumns{6}
\tablecaption{Summary of Observations\label{tab:obs}}
\tablehead{
        \colhead{UT Date}   &  \colhead{Target}  &  \colhead{ }    &    \colhead{}  &    \colhead{No. of}  &    \colhead{Coadds $\times$ }  \\
        \colhead{(Y/M/D)}   &   \colhead{(A/B)}  & \colhead{Instrument}    &    \colhead{Filter/$\lambda \lambda$}  &    \colhead{Exposures}  &    \colhead{Exp. Time (s)}  
        }   
\startdata
\multicolumn{6}{c}{Imaging} \\
\tableline
2011/06/21  &  A  &   Keck-II/NIRC2   & $H$  & 5 & 100$\times$0.028    \\
2011/06/21  &  A+B  &   Keck-II/NIRC2   & $H$  & 5 & 1$\times$10    \\
2011/11/15  &  A+B  &   Keck-II/NIRC2   & $K'$  & 9 & 50$\times$0.2      \\
2011/12/02 & A+B  &   IRTF/SpeX  & $Y$ & 8 & 2$\times$10  \\
2011/12/02 &  A+B  &   IRTF/SpeX  & $J$ & 20 & 5$\times$1.5  \\
2011/12/02 &  A+B &    IRTF/SpeX  & $H$ & 20 & 10$\times$1  \\
2011/12/02 &  A+B  &   IRTF/SpeX  & $K$ & 20 & 2$\times$2  \\

\tableline
\multicolumn{6}{c}{Spectroscopy} \\
\tableline

2011/10/14 &   B &   IRTF/SpeX-prism  & 0.8--2.5~$\mu$m & 6 & 1$\times$120    \\
2011/12/02 & A+B  &    IRTF/SpeX-SXD  & 1.1--2.5~$\mu$m & 24 & 1$\times$120     \\
2011/12/26 &  B  &   Keck-II/OSIRIS  & 1.18--1.35~$\mu$m & 6 & 1$\times$300

\enddata
\end{deluxetable}

\begin{deluxetable}{lcccccc}
\tabletypesize{\small}
\tablewidth{0pt}
\tablecolumns{7}
\tablecaption{Keck/NIRC2 Astrometry of 1RXS~J2351+3127~AB\label{tab:astrometry}}
\tablehead{
        \colhead{Epoch (UT)}   &  \colhead{Filter} &  \colhead{FWHM (mas)}  & \colhead{Strehl}  &  \colhead{Separation (mas)}    &   
         \colhead{PA ($^{\circ}$)}   &  \colhead{$\Delta$mag} 
        }   
\startdata
2011.470   &   $H$  &  39.2 ~$\pm$~0.5    &   0.435 $\pm$ 0.009  &   2392.2~$\pm$~2.0   &  91.77~$\pm$~0.05   &  5.68 $\pm$ 0.04  \\
2011.871   &   $K'$  &   49.0 $\pm$ 0.2   &  0.50 $\pm$ 0.02    &   2386.3~$\pm$~1.5   &  91.81~$\pm$~0.04   &  5.04 $\pm$ 0.05 
\enddata
\tablecomments{FWHM and Strehl ratios are computed using the publicly available IDL routine \texttt{NIRC2STREHL} made available by Keck Observatory.}
\end{deluxetable}

\begin{deluxetable}{lcc}
\tabletypesize{\small}
\tablewidth{0pt}
\tablecolumns{3}
\tablecaption{Photometry of 1RXS~J2351+3127~AB\label{tab:photometry}}
\tablehead{
        \colhead{Property}   &    \colhead{Primary}    &    \colhead{Secondary}
        }   
\startdata
$R_\mathrm{USNO-B}$ (mag)     &     12.28   &  $\cdots$   \\
$I_\mathrm{USNO-B}$ (mag)     &     10.92   &  $\cdots$   \\
$J_\mathrm{MKO}$ (mag)     &     9.80 $\pm$ 0.02\tablenotemark{a} &  $\cdots$  \\
$H_\mathrm{MKO}$ (mag)     &     9.21 $\pm$ 0.02\tablenotemark{a}  &  14.89~$\pm$~0.04\tablenotemark{b} \\
$K_\mathrm{MKO}$ (mag)     &     8.98 $\pm$ 0.02\tablenotemark{a}  & 13.92~$\pm$~0.05\tablenotemark{b}  \\
$M_{J\mathrm{(MKO)}}$ (mag)     &     6.31 $\pm$ 0.46\tablenotemark{c} &  $\cdots$ \\
$M_{H\mathrm{(MKO)}}$ (mag)     &   5.72  $\pm$ 0.46\tablenotemark{c}  &11.40~$\pm$0.46  \\
$M_{K\mathrm{(MKO)}}$ (mag)     &     5.49 $\pm$ 0.46\tablenotemark{c}  & 10.43~$\pm$0.46  \\
$\Delta Y$, IRTF (mag)               &     \multicolumn{2}{c}{5.71 $\pm$ 0.19}   \\
$\Delta J$,  IRTF (mag)               &     \multicolumn{2}{c}{5.27 $\pm$ 0.27}   \\
$\Delta H$, IRTF (mag)               &     \multicolumn{2}{c}{5.10 $\pm$ 0.22}   \\
$\Delta K_S$, IRTF (mag)               &     \multicolumn{2}{c}{4.54 $\pm$ 0.12}   \\
$GALEX$ $NUV$ (mag)    &     \multicolumn{2}{c}{19.97 $\pm$ 0.09\tablenotemark{d}}    \\
$GALEX$ $FUV$ (mag)    &     \multicolumn{2}{c}{21.29 $\pm$ 0.26\tablenotemark{d}}    \\
$ROSAT$ flux (erg sec$^{-1}$ cm$^{-2}$)    &     \multicolumn{2}{c}{6.4 $\pm$ 1.5 $\times$10$^{-13}$\tablenotemark{c}}    \\
$ROSAT$ HR1    &     \multicolumn{2}{c}{--0.33 $\pm$ 0.16\tablenotemark{c}}    \\
$ROSAT$ HR2   &     \multicolumn{2}{c}{--0.03 $\pm$ 0.30\tablenotemark{c}}    

\enddata

\tablenotetext{a}{Synthetic photometry from our SXD spectrum of 1RXS~J2351+3127~A after 
  flux-calibrating it to the 2MASS $K_S$-band magnitude from \citet{Skrutskie:2006p589}.}
  \tablenotetext{b}{Computed from our Keck/NIRC2 photometry.}
\tablenotetext{c}{Based on the photometric distance estimate of 50~$\pm$~10~pc.}
\tablenotetext{d}{$GALEX$ photometry from GR6 (\citealt{Morrissey:2007p22251}).}
\tablenotetext{e}{From the $ROSAT$ All-Sky Survey (\citealt{Voges:1999p22945}).  
The relation from \citet{Fleming:1995p23307} was used to convert count rate to flux.}
\end{deluxetable}

\begin{deluxetable}{lcc}
\tabletypesize{\small}
\tablewidth{0pt}
\tablecolumns{3}
\tablecaption{Properties of 1RXS~J2351+3127~AB\label{tab:properties}}
\tablehead{
        \colhead{Property}   &    \colhead{Primary}    &    \colhead{Secondary}
        }   
\startdata
Age (Myr)     &     \multicolumn{2}{c}{50--150\tablenotemark{a}}   \\
$d_\mathrm{phot}$ (pc)     &     \multicolumn{2}{c}{50~$\pm$~10}   \\
Proj. Sep. ($''$)     &      \multicolumn{2}{c}{2.386 $\pm$ 0.002}   \\
Proj. Sep. (AU)     &     \multicolumn{2}{c}{119 $\pm$ 24}  \\
$\mu_{\alpha}$cos $\delta$ (mas/yr)    &     \multicolumn{2}{c}{105.9 $\pm$ 3.5\tablenotemark{b}}   \\
$\mu_{\delta}$ (mas/yr)    &    \multicolumn{2}{c}{--81.8 $\pm$ 5.3\tablenotemark{b}}   \\
$RV$ (km/s)    &   \multicolumn{2}{c}{--13.5 $\pm$~0.6\tablenotemark{c}}   \\
$U$ (km/s)    &   \multicolumn{2}{c}{--10.5 $\pm$ 3.0}   \\
$V$ (km/s)    &   \multicolumn{2}{c}{--29.2 $\pm$ 3.7}   \\
$W$ (km/s)    &   \multicolumn{2}{c}{--15.4 $\pm$ 4.6}   \\
$X$ (pc)    &   \multicolumn{2}{c}{--13.6 $\pm$ 2.7}   \\
$Y$ (pc)    &   \multicolumn{2}{c}{--41.2 $\pm$ 8.3}   \\
$Z$ (pc)    &   \multicolumn{2}{c}{--24.8 $\pm$ 5.0}   \\
log($L_\mathrm{X}$/$L_\mathrm{Bol}$)    &     \multicolumn{2}{c}{--3.02\tablenotemark{d}}   \\
log($L_\mathrm{Bol}$/$L_{\odot}$)    &     --1.37~$\pm$~0.19   &  --3.6~$\pm$~0.2   \\
Spectral Type    &     M2.0$\pm$0.5\tablenotemark{e}   &  L0$^{+2}_{-1}$   \\
Mass     &     0.45~$\pm$~0.05~\Msun   &  32~$\pm$~6~\Mjup   
\enddata
\tablecomments{$UVWXYZ$ values are based on the photometric distance estimate.  $U$ is positive towards the galactic center.}
\tablenotetext{a}{Assumes the system is a member of the AB Dor YMG.}
\tablenotetext{b}{UCAC-3; \citet{Zacharias:2010p23124}.}
\tablenotetext{c}{Shkolnik et al. (2012, submitted).}
\tablenotetext{d}{\citet{Riaz:2006p20030}.}
\tablenotetext{e}{\citet{Shkolnik:2009p19565}.}
\end{deluxetable}

\end{document}